\documentclass{acm_proc_article-sp}
\usepackage{multirow}
\usepackage{balance}
\usepackage{hyperref}
\usepackage{graphicx}
\usepackage{xspace}
\usepackage{subcaption}
\usepackage{color}
\usepackage{tikz}
\usepackage{listings}
\usepackage{epstopdf}
\usepackage[lined,ruled,linesnumbered]{algorithm2e}
\usepackage{booktabs}
\usepackage{subcaption}
\usepackage{lipsum}
\usepackage{array}

\newtheorem{mydef}{Definition}

\newcommand{\squishlist}{
  \begin{list}{$\bullet$}
   {
     \setlength{\itemsep}{0pt}
     \setlength{\parsep}{0pt}
     \setlength{\topsep}{0pt}
     \setlength{\partopsep}{0pt}
     \setlength{\leftmargin}{1.5em}
     \setlength{\labelwidth}{1em}
     \setlength{\labelsep}{0.5em} } }
\newcommand{\squishend}{
   \end{list}  }
   
\newcommand{\system}{Sapphire\xspace}

\definecolor{annotationColor}{RGB}{0,0,139}

\newcommand{\hide}[1]{}

\begin{document}

\title{Sapphire: Querying RDF Data Made Simple}

\numberofauthors{4} 
%
\author{
\alignauthor
Ahmed El-Roby\\
       \affaddr{Carleton University}\\
       \email{\small{ahmed.elroby@carleton.ca}}
\alignauthor
Khaled Ammar\\
       \affaddr{University of Waterloo}\\
       \email{\small{kammar@uwaterloo.ca}}
\alignauthor Ashraf Aboulnaga\\
       \affaddr{Qatar Computing Research Institute - HBKU}\\
       \email{\small{aaboulnaga@qf.org.qa}}
\alignauthor
Jimmy Lin\\
       \affaddr{University of Waterloo}\\
       \email{\small{jimmylin@uwaterloo.ca}}
}

\maketitle

\begin{abstract}

\begin{sloppypar}
RDF data in the linked open data (LOD) cloud is very valuable for many different applications. In order to unlock the full value of this data, users should be able to issue complex queries on the RDF datasets in the LOD cloud. SPARQL can express such complex queries, but constructing SPARQL queries can be a challenge to users since it requires knowing the structure and vocabulary of the datasets being queried. In this paper, we introduce \system, a tool that helps users write syntactically and semantically correct SPARQL queries without prior knowledge of the queried datasets. \system interactively helps the user while typing the query by providing auto-complete suggestions based on the queried data. After a query is issued, \system provides suggestions on ways to change the query to better match the needs of the user. We evaluated \system based on performance experiments and a user study and showed it to be superior to competing approaches.
\end{sloppypar}

\end{abstract}

\section{Introduction}
\label{sec:introduction}

In recent years, advances in the field of information extraction have helped in automating the construction of large Resource Description Framework (RDF) datasets that are published on the web. These datasets could be general-purpose such as DBpedia\footnote{\url{http://dbpedia.org}}~\cite{Auer2007},
a dataset of structured information extracted from Wikipedia,
or they could be specific to particular domains such as
movies\footnote{\url{http://www.linkedmdb.org}},
geographic information\footnote{\url{http://www.geonames.org}},
and city data\footnote{\url{http://www.data.gov/opendatasites}}.
These datasets are graph-structured, and 
are interlinked
 via
edges that point from one dataset to another, forming a massive graph known as the \textit{Linked Open Data (LOD) cloud}\footnote{\url{http://lod-cloud.net}}.

The LOD cloud contains a wealth 
of structured information that can be
extremely useful to users and applications in diverse domains. However,
utilizing this information
requires an effective way to find the
answers to questions in 
the datasets that make up this cloud. 
Answering questions over RDF data generally follows one of two approaches:
(a) natural language queries, and (b) structured querying using SPARQL~\cite{SPARQL}, the standard query language for RDF. 


Natural language approaches rely on keyword search or more complex question answering techniques.
These approaches are convenient and easy to use, and they find accurate
answers for simple questions such as \emph{``How many people live in
  New York?''}.
Questions like this one that ask about a specific property of an entity are
termed \textit{factoid questions}.
Such questions can be answered by a simple structured search
that can be constructed effectively by natural language approaches.

However, the RDF data
that makes up the LOD cloud
is
not limited to answering simple questions.
This data can be used to answer complex questions
that require complex structured searches.
Natural language approaches are not effective at constructing
such complex structured searches.
Instead, complex structured searches are better expressed 
using SPARQL queries.
It is a common practice for data sources in the LOD cloud to provide
\textit{SPARQL endpoints} that allow users 
to issue
SPARQL queries on the RDF data that they contain\footnote{e.g., \url{http://dbpedia.org/sparql} for DBpedia.}.

\begin{sloppypar}
To illustrate the need for SPARQL, consider the question
\emph{``How many scientists graduated from an Ivy League university?''}
This question was used in the QALD-5 competition~\cite{Unger2015}.
QALD is an annual competition for Question Answering over Linked Data,
and this question was not answered by any of the natural language systems
that participated in QALD-5.
This is not surprising since the question involves concepts such as
``scientist'', ``graduated'', and ``Ivy League university'' that are not easy to map
to a structured search over the queried dataset (DBpedia), in addition to requiring a count of the results.
On the other hand,
the following query over the SPARQL endpoint of DBpedia will find the
required answer:\end{sloppypar}
\begin{verbatim}
PREFIX res: <http://dbpedia.org/resource/>
PREFIX dbo: <http://dbpedia.org/ontology/> 
SELECT DISTINCT count (?uri) WHERE {
 ?uri rdf:type dbo:Scientist.
 ?uri dbo:almaMater ?university.
 ?university dbo:affiliation res:Ivy_League.
}\end{verbatim}
To be able to
compose a query such as this one,
the user needs to know the structure of the dataset,
the vocabulary used to represent different concepts,
and the literals used in the dataset including their data types and format.
For example, the user needs to know that ``scientist'' is an
\texttt{rdf:type} 
and ``Ivy League'' is an \texttt{affiliation} of a university.
Achieving this level of knowledge about a dataset
can be difficult
even for experienced users given the
massive scale and diverse vocabulary of the LOD cloud.
By one recent count\footnote{\url{http://stats.lod2.eu}}, the LOD cloud has almost 3000 data sources that contain over 14 billion RDF triples from various domains.
To illustrate the diversity of the vocabulary in the LOD cloud,
consider that DBpedia alone has over 3K distinct predicates at the time of writing this paper.
Thus, it is quite likely that a user would need to construct SPARQL queries on data whose structure and vocabulary she does not know in full,
for example when querying a new dataset.
Our goal in this paper is to help users with this challenging task.

We present \textit{\system},
an interactive tool aimed at helping users write
syntactically and semantically correct SPARQL queries
on RDF datasets they do not have
prior knowledge about.
\system is aimed at users who have a technical background but are not necessarily SPARQL experts, e.g.,
data scientists or application developers.
Thus, \system makes no attempt to ``shield'' its users from
the syntax of SPARQL, but rather helps them construct valid SPARQL queries with ease.
\system achieves this in two ways that both rely on a \emph{predictive user model}
that is built for the endpoints to be queried in an initialization phase.
First, while a user is typing a query, 
\system interactively
provides her with data-driven suggestions to complete the predicates and literals
in the query,
similar to the auto-complete capability in many user interfaces.
Second, when a user completes the query and submits it for execution,
\system suggests ways to modify the query
into one that may be better suited to the needs of the user.
For example, if the user 
query
returns no answers, \system would attempt
to modify it into a query that does return answers.


\system's query completion and query suggestion modules
rely on  natural language techniques.
Thus, in the spectrum of approaches for querying RDF, \system bridges the gap between the simple but ambiguous natural language approaches on the one hand, and the powerful but
cumbersome SPARQL on the other.
The novelty of \system comes from the need
to balance multiple conflicting goals: 
\system must provide high quality recommendations that actually help the user find
the information that she needs, 
it must have fast response time since it is interactive, 
it must run on a reasonably sized machine without 
placing excessive demands on the
machine's resources,
and it must not overload the SPARQL endpoints it queries.
These design goals require judicious design choices which we
present in the rest of this paper.
We have built \system as a web based querying
tool\footnote{\url{http://github.com/aelroby/Sapphire}},
and we demonstrated its user interface and query composition 
workflow in~\cite{El-Roby2016}.
In this paper, we present the internals of \system and demonstrate
through experiments and a user study that it is significantly more effective
than competing approaches in finding answers to user queries,
and it achieves interactive performance.


We review related work in Section~\ref{sec:related} and present
the architecture of \system in Section~\ref{sec:architecture}. We present the Sapphire user interface from~\cite{El-Roby2016} in Section 4.
We then present the contributions of the paper,
 which are as follows:
\squishlist
\item Summarizing the queried endpoints to collect concise, important data that is utilized by \system (Section~\ref{sec:initialization}).
\item The predictive user model which is at the heart of \system and includes two modules: query completion and query suggestion (Section~\ref{sec:pum}).
\item An extensive evaluation of \system based on performance experiments and a user study (Section~\ref{sec:evaluation}).
\squishend

\section{Related Work}
\label{sec:related}
\begin{figure*}
	\centering
	\includegraphics[height=5.5cm,width=16.5cm]{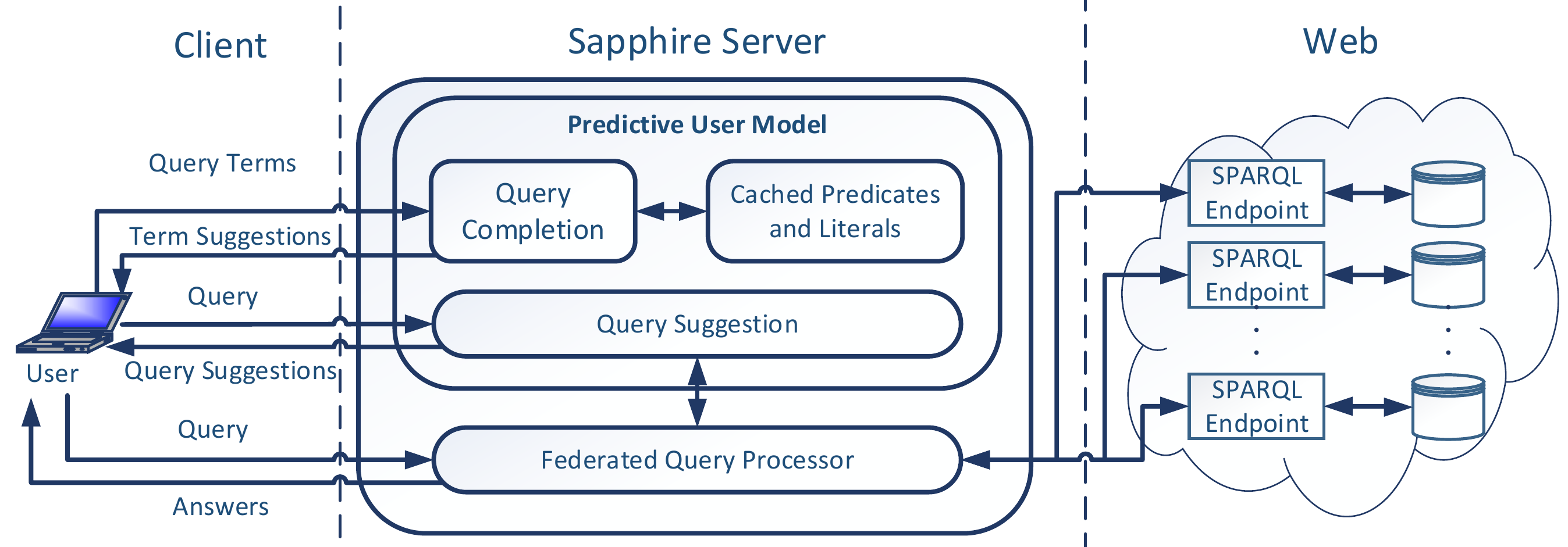}
	\caption{Architecture of \system.}%
	\label{fig:architecture}%
\end{figure*}
Prior work on helping users construct structured queries on RDF data
falls into three categories:
1.~Natural language approaches.
2.~Approximate 
 queries.
3.~Query by example.

\begin{sloppypar}
\textbf{Natural Language Approaches}: Several prior works create structured queries based on natural language approaches~\cite{Dong2007,Lopez2011,Unger2012,Yahya2013,Fader2014}.
Each of these works focuses on one or more specific 
query templates,
and 
uses keyword search or natural language questions to 
construct these templates and 
fill in the placeholders they contain.
All of these approaches suffer from two limitations compared to \system:
(1)~Their expressiveness is limited
to specific query templates, and (2)~inferring query structure, predicates, and literals based only on
natural language is inherently ambiguous.
In contrast, \system  can construct any SPARQL query,
and it removes ambiguity by involving the user directly in
query composition.
\end{sloppypar}

In this paper, we compare \system to QAKiS~\cite{Cabrio2012} and
KBQA~\cite{Cui2017} as representatives of the state of the art in
natural language approaches,
and we show that \system outperforms these two systems.
QAKiS~\cite{Cabrio2012} is a question answering system over RDF that
automatically extracts from Wikipedia different ways of expressing relations
in natural language (e.g., ``a bridge spans a river'' and ``a bridge crosses a river'' express the same relation).
These equivalent expressions are used to match fragments of a natural language question and 
construct the equivalent SPARQL query.
KBQA~\cite{Cui2017} is a more recent question answering system that focuses on
factoid questions.
KBQA learns question templates from a large Q\&A corpus
(e.g., Yahoo! Answers),
and learns mappings from these templates to RDF predicates
in the queried dataset.
The templates and mappings are then used to answer user questions.

\textbf{Approximate Queries}: This line of work goes beyond the 
fixed query templates used by natural language approaches, 
enabling the user to express approximate structured queries.
That is, the query posed by the user does not have to be exactly
matched with the queried RDF data~\cite{Kochut2007,Kiefer2007,Yang2014}.
These approaches are still limited in the query structure that they
support,
and they require the user to know the vocabulary and the approximate schema of the queried datasets.
In contrast, \system enables the user to compose any SPARQL query 
without prior knowledge of the queried datasets. 

We compare \system to $S^{4}$~\cite{Zheng2016},
a recent system that was shown to outperform other approximate query
approaches.
$S^{4}$ summarizes 
the queried dataset
by
maintaining a graph of the relationships between RDF entity types based on
the relationships between instances of these types.
Queries are rewritten based on this graph.
$S^{4}$ assumes that the user can issue queries using correct 
predicates and instance URIs in the dataset,
but possibly not with the correct query structure.

\textbf{Query by Example}:
SPARQLByE~\cite{Arenas2016,Diaz2016}
 infers the SPARQL query that best suits the user's needs based on a set of example answers she provides.
A key limitation of this approach is that the user needs to know a set of examples that satisfy her query,
which is often not practical.
For example, to answer the query
\emph{``How many people live in New York?''}, the user
should know the precise population of some cities to provide as examples, which can be impractical.
In contrast, \system helps the user directly construct
a SPARQL query rather than
inferring it.
We compare \system to SPARQLByE and show that \system is more expressive.



\section{\system Architecture and Challenges}
\label{sec:architecture}

In this section we present the overall architecture of Sapphire, and an overview of the different design choices and challenges that must be addressed in order to implement a useful and efficient system.

Figure~\ref{fig:architecture} shows the architecture of \system, which runs
as a server that sits between the user and the SPARQL endpoints for one or more RDF datasets on the web. \system accesses the endpoints through a federated query processor. \system uses FedX~\cite{Schwarte2011}, a widely-used federated query processor,
but any other federated query processor can be used.

The core of \system is the 
Predictive User Model (PUM), which helps the user express her information needs using
SPARQL queries.
The PUM relies on information about the datasets being queried. Before querying an endpoint, the user must register this endpoint with the \system server, and the server goes through an initialization step in which it caches important data from this endpoint.
One challenge that must be addressed by \system is which data from an endpoint to cache,
and how to retrieve this data without overloading the endpoint.

While the user is composing a query, the query terms are forwarded to the Query Completion Module (QCM) as they are typed by the user. The QCM interactively provides the user with suggestions to complete the terms in her query
based
on the data cached during initialization.
A question that must be answered when designing the QCM is how to provide interactive response time even for the large scale data 
in the LOD cloud.

After composing a syntactically correct query, the federated query processor executes the query and returns answers.
Simultaneously, the Query Suggestion Module (QSM) suggests changes to the query to help the user find the answers she is looking for. The goal of the QSM is to suggest queries that are similar to the one issued by the user, but different enough to present her with useful alternatives that may help her satisfy her information needs. These suggestions span two directions: 1.~Finding alternative literals and predicates to the ones used in the query. 
2.~Relaxing the structure of the issued query to \textit{approximately}
match the issued query with 
candidate patterns in the dataset.
Query suggestions are provided for all queries,
and
it is up to 
the user to accept these suggestions if the returned answers do not satisfy her information needs.
The QSM poses several interesting research questions, such as which literals and predicates to replace in the query
and how to find replacement terms. Also, what does it mean to relax the structure of a query and how to find the relaxed
structure efficiently. The way we address the different requirements and challenges in Sapphire is described in the next three sections. We start by discussing the user interface in Section~\ref{sec:UI}, then we present how initialization happens for a new endpoint in Section~\ref{sec:initialization}, and the PUM in Section~\ref{sec:pum}.

\section{User Interface}
\label{sec:UI}

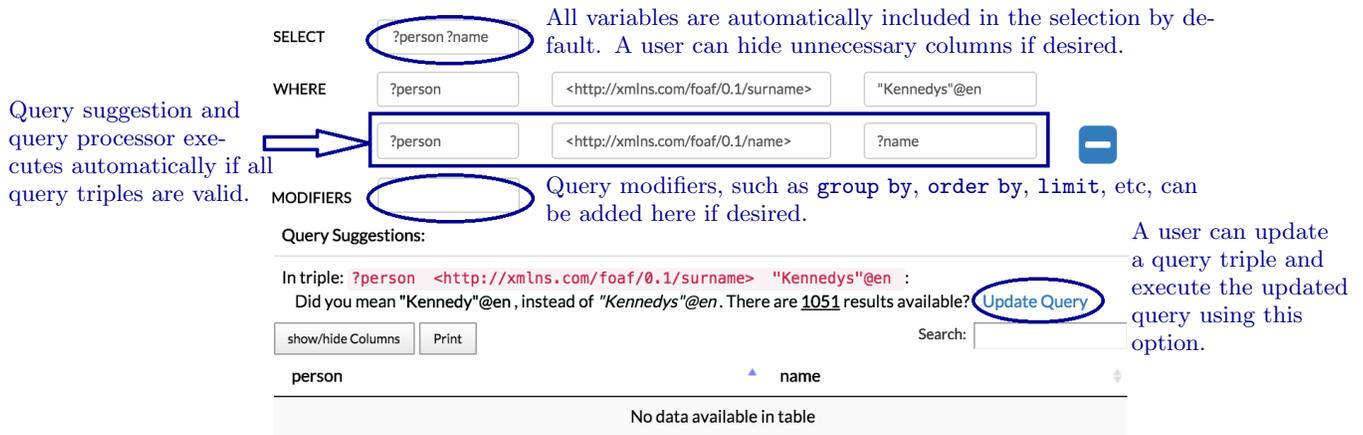
\begin{figure*}[t]
	\centering

\begin{tikzpicture}
\draw (0, 0) node[inner sep=0] {\includegraphics[width=18cm]{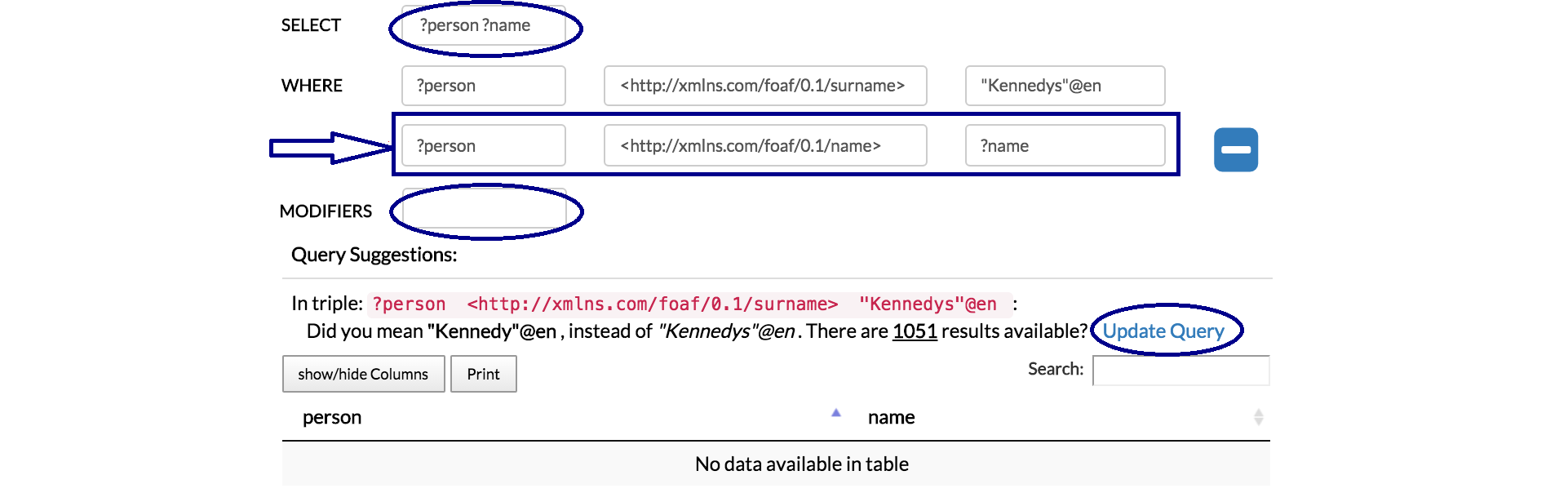}};
\draw (-7.5, 1) node[text width=11em] {\color{annotationColor}Query suggestion and query processor executes automatically if all query triples are valid.};
\draw (2.4, 2.6) node[text width=28em] {\color{annotationColor}All variables are automatically included in the selection by default. A user can hide unnecessary columns if desired.};
\draw (7.1, -0.8) node[text width=9em] {\color{annotationColor}A user can update a query triple and execute the updated query using this option.};
\draw (2.4, 0.4) node[text width=28em] {\color{annotationColor}Query modifiers, such as \texttt{group by}, \texttt{order by}, \texttt{limit}, etc, can be added here if desired.};
\end{tikzpicture}
	\caption{User interface showing a suggestion to modify the current query which returned to answers.}%
	\label{fig:alternatives}%
\end{figure*}

\begin{figure}
\centering
\includegraphics[height=4cm,width=4cm]{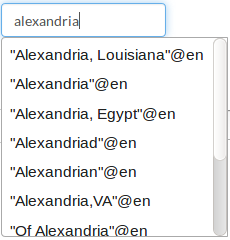}
\caption{Auto-complete suggestions using the QCM.}
\label{fig:typeahead}
\end{figure}

\begin{figure*}[t]
	\centering
	\begin{tikzpicture}
	\draw (0, 0) node[inner sep=0] {\includegraphics[width=17cm]{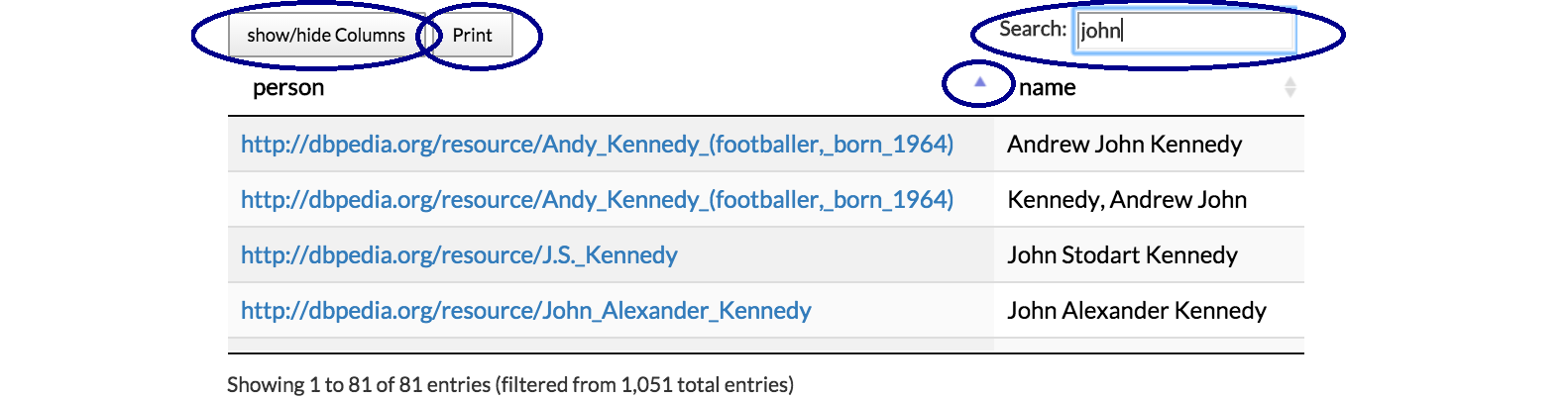}};
	\draw (-8, 2) node[text width=10em] {\color{annotationColor}Controls the visibility of columns.};
	\draw (-0.1, 1.8) node[text width=15em] {\color{annotationColor}Prepare a printable version.};
	\draw (7.2, 0.8) node[text width=9em] {\color{annotationColor}Search capability allows users to filter results using keyword search.};
	\draw (0, 1.2) node[text width=15em] {\color{annotationColor}Sort answers by any column.};
	\end{tikzpicture}
	
	\caption{The answer table after applying the query suggestion in Figure~\ref{fig:alternatives}. In this example, the 1,051 answers to the query are filtered via a keyword search on ``\texttt{john}'', and the filtered answers are ordered by the ``\texttt{person}'' column.}%
	\label{fig:answers}%
\end{figure*}

\begin{sloppypar}
Sapphire has a web-based user interface that was demonstrated in~\cite{El-Roby2016}.  This interface is shown in Figure~\ref{fig:alternatives}. The interface presents a text box for each part of a SPARQL query. While the user is typing query terms, the QCM provides suggestions to complete these terms as shown in Figure~\ref{fig:typeahead}. After the user inputs a query, the query is validated and executed. Whenever a query is executed, the QSM tries to find alternatives to the query that was constructed by the user. 
Figure~\ref{fig:alternatives} shows an example of how the QSM suggests changes to the executed query. In this example, the user wants to find all people with the surname ``Kennedys'' (in plural form). However, no answers were found using this surname. The QSM suggests a modification that will result in finding 1,051 answers, by changing ``Kennedys'' to ``Kennedy''. If the user accepts this suggestion and updates the query, the new query is executed and the answers are displayed in the answer table (Figure~\ref{fig:answers}). New suggestions are now displayed to the user in case these answers still do not satisfy her information needs. The query alternatives are shown to the user in the form of suggestions to change one term at a time. For example, one suggestion could be ``In the triple \texttt{(subject, predicate, object)}, did you mean \texttt{predicate}, instead of \texttt{predicate}? There are $N$ answers available.''. This approach avoids showing the user a completely rewritten SPARQL query in one step, which would make the suggestions difficult to understand, especially for large and complex queries. The only exception is when the QSM suggests queries that are different in structure than the issued query. We will elaborate on this specific type of query suggestion in Section~\ref{subsec:qsm}.
\end{sloppypar}

\begin{sloppypar}
The suggested queries are executed in the background using the Federated Query Processor and answers are prefetched so that when the user decides to choose one of the alternatives, the query is not re-executed, and the answers are displayed almost-instantaneously. When the answers to a query are displayed to the user, she has the ability to manipulate them in the answer table, as shown in Figure~\ref{fig:answers}. Supported operations include the following:\ the user can search the answer table using a keyword search box, order the answers by any column, show and hide columns, and drag and drop answers from the answer table to the query text boxes for additional queries. Next, we turn our attention to the technical details of Sapphire initialization and the PUM.
\end{sloppypar}
\section{Initialization for a New Endpoint}
\label{sec:initialization}


This section describes the initialization step in which \system retrieves data from
a newly registered endpoint. That is: 1.~Which data from the endpoint to cache? 2.~How is this data retrieved? 3.~How is it indexed for efficient access by the PUM?

\subsection{Data Caching}
\label{subsec:cachedData}

The data cached by \system from the endpoints
plays a significant role in helping the user write a query that describes her information needs.
In designing \system,
we assume that it is simpler and more intuitive for users to express their information
needs using keywords rather than using URIs.
Therefore, the focus of the \system PUM is on mapping keywords entered by the user
in her query to RDF predicates and literals in the dataset.

Thus, \system needs to cache RDF predicates and literals from a dataset
so that these predicates and literals can
subsequently be
matched to keywords
in the user query.
Which predicates and literals to cache is a challenging question.
The choice of data to cache cannot rely on statistical knowledge of the queried datasets or the query logs, since such knowledge is not available.
Therefore, we develop heuristics based on common characteristics of RDF datasets and SPARQL queries.

Our first heuristic relies on the observation that 
the number of distinct predicates in a dataset is typically much smaller than the number of distinct literals. For example, at the time of this writing,
DBpedia has approximately 3K distinct predicates compared to 70M distinct literals. Therefore, \system caches all the predicates in a dataset.

Given the typically large number of literals in a dataset, \system uses heuristics to limit the number of literals that it caches.
First, we assume that very long literals are not likely to be used in queries.
Thus, \system only caches literals below a certain length (in this paper we use 80 characters as the limit).
Second, we assume that the user
is
interested only in 
a certain language
and
allow the user to restrict the language of the cached literals (in this paper we cache only English literals).

Following the aforementioned heuristics reduces the number of cached literals. However, the number of literals that satisfy these heuristics will likely be too large to retrieve from the endpoint using a single SPARQL query.
Such a query would be a long-running query, and most endpoints impose a timeout limit on 
queries
to avoid overloading their computing resources, or reject queries from the start if their estimated execution time is above a
threshold.
Thus, we need to decompose this query into multiple queries that are each within the timeout limit. 
Furthermore, we need to ensure that the entire initialization process finishes within a reasonable amount of time.
Our goal is for initialization time to be on the order of hours, which we believe is reasonable since the initialization process happens only once for each endpoint.
Next, we describe the queries that we use to retrieve literals from an endpoint for caching.

\system divides the dataset based on the predicates and the class hierarchy defined by RDF schema (RDFS)~\cite{RDFS}.
RDFS defines classes that serve as data types for different entities,
and organizes the classes into a hierarchy based on the \texttt{subClassOf} relation. For example, \texttt{MovieDirector} and \texttt{Politician} are two classes that are both subclasses of \texttt{Person}.
\system issues a SPARQL query to retrieve all classes and their subclasses from the endpoint. It also issues a query to retrieve all RDF predicates associated with literals, ordered by the numbers of literals associated with each predicate. These are short queries that are not expected to time out.
\system then iterates through the predicates associated with literals, from most frequent to least frequent.
For each predicate, \system navigates through the class hierarchy from root to leaves.
At each class of the hierarchy, \system creates a query to retrieve literals associated with the current predicate and current class, and that are below the threshold length and in the target language.
To increase the likelihood that this query will succeed, it  is decomposed into multiple queries using SPARQL pagination techniques (OFFSET and LIMIT).
If this query succeeds, \system moves to the next sibling in the class hierarchy.
If this query times out, \system navigates down to the next level of the class hierarchy, which contains smaller classes, and issues the query. 
This process continues until all the literals are retrieved.
\system allows the user to set a limit on the 
number of queries to issue to an endpoint
and
stops when this limit is reached.
Since   
\system orders predicates by frequency,
it prioritizes caching the literals associated with frequent predicates.
 
For the uncommon case of datasets that do not use the class hierarchy of RDFS (about 75\% of the datasets in the LOD cloud
use RDFS\footnote{\url{http://stats.lod2.eu}}), \system issues a query to retrieve the entity types that occur frequently in the dataset.
\system then issues queries to retrieve the literals associated with each predicate and each of these entity types, iterating through the predicates and types from most frequent to least frequent.
If there is a limit on the number of queries, \system stops if this limit is reached.
The complete list of queries that are sent to an endpoint during initialization is presented in Appendix~\ref{app:bootstrappingQueries}.

\subsection{Indexing Cached Data}
\label{subsec:indexing}

As discussed earlier, one of the key challenges facing \system is
providing suggestions to the user interactively.
These suggestions come from the cached data, so 
this data must be indexed in a way that supports fast lookup.

The basic lookup operation for suggesting completions to the user is as follows:
given a string $t$ entered by the user, what strings in the data contain $t$?
We observe that a \textit{suffix tree}~\cite{Weiner1973} is ideally suited for
this type of lookup, so we use it as an index in \system.
The advantage of a suffix tree is that lookup operations depend only on the size of the lookup string $t$ and the number of times $z$
that this
string occurs in the input,
with a 
time complexity $O(|t| + z)$.
The disadvantage of a suffix tree is that it can grow very large, sometimes over an order of magnitude larger than the size of the input.

Given the space consumption of suffix trees, only a subset of the cached data can be indexed in this tree.
Since the number of RDF predicates is relatively small, all predicates are indexed. The more challenging question is which subset of the literals to index?
To answer this question, we introduce the notion of
\textit{most significant} literals, and index only these literals in the suffix tree. A literal is considered significant when the entity it is associated with occurs frequently in the dataset. That is, there are many incoming edges in the RDF graph pointing to this entity, indicating the entity's importance.

\begin{mydef}
The significance score of a literal $l$ is $S(l) = |\{ s | (s, p_{1}, o) \land (o, p_{2}, l)\}|$,
where $(s, p_{i}, o)$ is an RDF triple.
\end{mydef}

 
For example, the literal ``New York'' is associated with the entity representing this city.
Since
this entity
is pointed to by many other entities (i.e., occurs as an RDF object), the literal ``New York'' is significant.
This definition of significance captures important classes in the RDF class hierarchy,
and also captures important instances (people, locations, etc.).
To identify the significant literals, \system issues queries along the class hierarchy as it did for retrieving the literals.

The final issue related to initialization is how to lookup in cached literals not in the suffix tree. We call these the \textit{residual literals}.
Lookup on the residual literals requires a sequential search,
and we have found that this may be too slow for interactive response.
To speed up this sequential search, \system organizes the literals into
\textit{bins of residual literals}, or \textit{residual bins} for short,
where each bin has all the literals of a given length
(i.e., $bin(literal) = |literal|$).
As discussed in the next section,
the PUM always searches for strings within a certain range of lengths,
so its sequential search will be limited to a few bins. In addition, the search can be parallelized,
with multiple threads simultaneously scanning the bins.
We show in Section~\ref{sec:evaluation} that this simple organization
is effective at guaranteeing interactive performance.

To illustrate the cost of initialization, we note that 
initialization for DBpedia, one of the largest datasets in the LOD cloud, took 17 hours.
In the process, \system issued approximately 
800 SPARQL queries to retrieve literals and 3000 to identify significant literals, in addition to the few queries that retrieve predicates  and the class hierarchy. Approximately 200 queries timed out.
The suffix tree for DBpedia contains 43K strings (3K predicates and 40K 
literals) and is 400MB in size.
There are 
around 
21M literals not in the suffix tree, divided among 80 bins.
We show in Section~\ref{sec:evaluation} that having even a small fraction of the literals in the suffix tree
benefits performance.

\section{Predictive User Model}
\label{sec:pum}

\begin{figure}
\centering
\includegraphics[height=7cm,width=8cm]{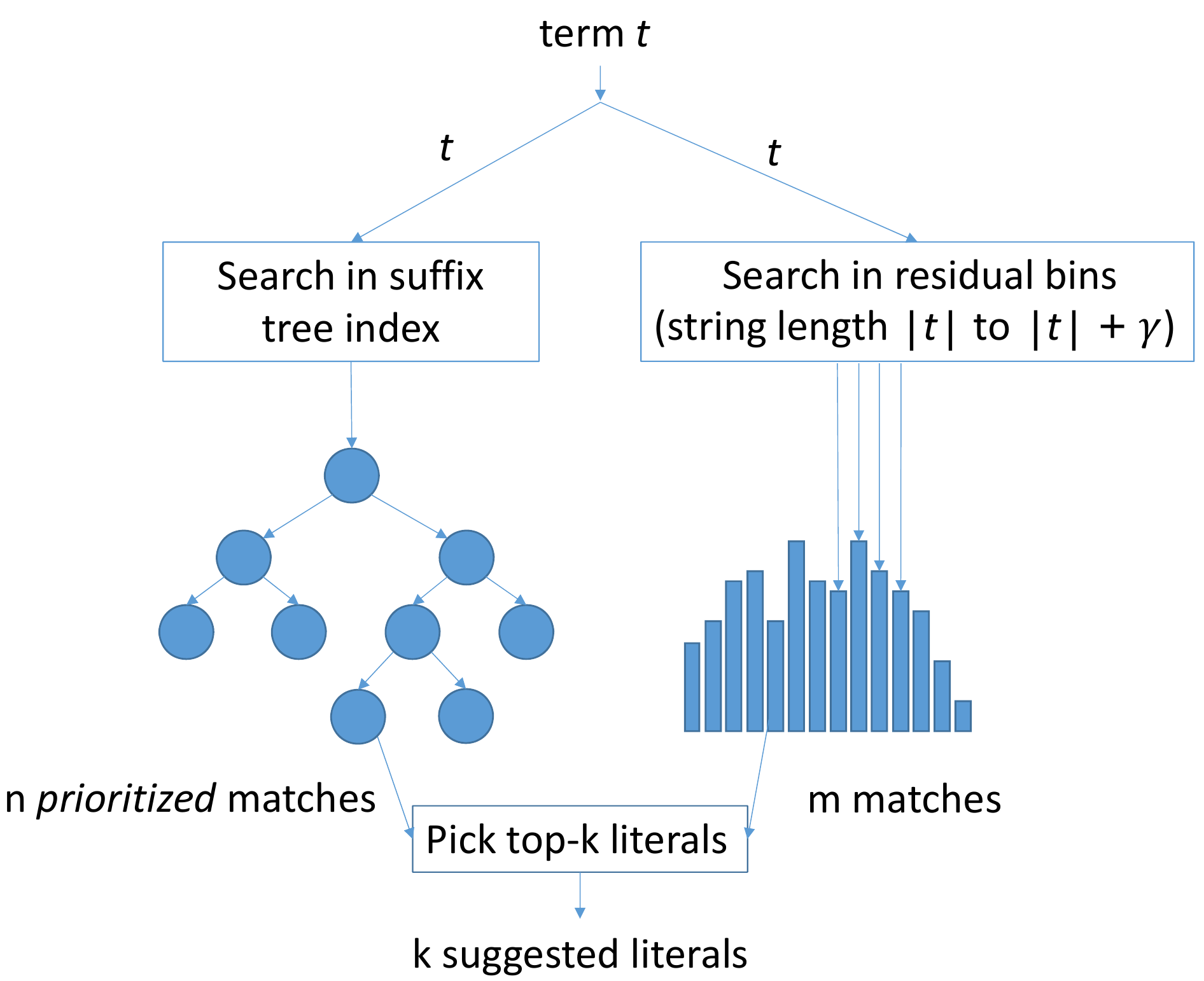}
\caption{Completing a query term in the QCM.}
\label{fig:qcmFramework}
\end{figure}

\begin{sloppypar}
The Predictive User Model (PUM) uses the data cached during initialization
to help the user compose her SPARQL query.
The user inputs a query to \system by entering the triple patterns
that describe the structure of the query.
As the user types a subject, predicate, or object in a triple pattern,
the PUM invokes the Query Completion Module (QCM) to provide suggestions
for the user to complete the term being typed.
When the user composes a full query and clicks ``Run'' in the \system user interface, the PUM passes the query to the federated
query processor
for execution and also invokes the Query Suggestion Module (QSM) to suggest
changes 
to the query.
The QSM suggests changes to the query based on the structure of potential candidate answers in the dataset,
in order to bring the user closer to the query that finds the answer she is
looking for.
The user can choose one of the suggestions of the QSM and update the query, and possibly continue editing it.
Editing the query would invoke the QCM again, and the process repeats as many
times as needed by the user.
We present the QCM next, followed by the QSM.
\end{sloppypar}


\subsection{Query Completion Module}
\label{subsec:qcm}

%
%
%
%

\begin{algorithm}[t]
\SetKwData{Left}{left}\SetKwData{This}{this}\SetKwData{Up}{up}
\SetKwFunction{Union}{Union}\SetKwFunction{FindCompress}{FindCompress}
\SetKwInOut{Input}{input}\SetKwInOut{Output}{output}

\Input{Bins to Search ${bins}'$, Number of Processes $P$}
\Output{Assigned Task for Each Process}
\BlankLine

Number of literals to search $n = \sum_{i = 1}^{|{bins}'|} |{bins}'_{i}|$\;

Process capacity $d = \frac{n}{P}$\;

Process id $pid = 0$\;
\For{$i = 1$ \KwTo $|{bins}'|$}{
Number of literals remaining in bin $i$ $j = |{bins}'_{i}|$\;
\While{$j > 0$}{
\eIf{$j < Process_{pid}.d$}{
// Process $pid$ assigned all literals in bin
$Assign (Process_{pid},$\\$ \left [{bins}'_{i}[0], {bins}'_{i}\left[|{bins}'_{i}|-1\right]\right ])$\;
$Process_{pid}.d = Process_{pid}.d - |{bins}'_{i}|$\;
$j = 0$\;
}{
// Process $pid$ assigned remaining capacity
$Assign(Process_{pid}, [{bins}'_{i}[|{bins}'_{i}| - j],$\\$ {bins}'_{i}[|{bins}'_{i}| - j + Process_{pid}.d]])$\;
$j = j - Process_{pid}.d$\;
$Process_{pid}.d = 0$\;
$pid = pid + 1$\;
}
}
}
\caption{Assign Tasks to Processes}
\label{algo:workload}
\end{algorithm}

The \system user interface is organized so that the user inputs each
subject, predicate, or object of a triple pattern in a separate text box, as shown in Figure~\ref{fig:alternatives}.
As the user types a string in one of these text boxes, the QCM is invoked
every time the user types a character in order to provide auto-complete
suggestions for the string being typed. The only exception is if the user enters a variable (i.e., a string starting with `?'), in which case \system makes no suggestions.

Specifically, the problem solved by the QCM is as follows:
Given the string $t$ entered thus far by the user, 
find $k$ strings in the data that contain $t$ to suggest to the user.
In this paper, we use $k = 10$.
Figure~\ref{fig:qcmFramework} shows how the QCM finds the required $k$ strings.
The term $t$ is looked up in both the suffix tree and the residual bins. Matches in the suffix tree are returned to the user as soon as they are
found. If the search in the suffix tree returns fewer than $k$ matches,
the remaining matches come from the residual bins.
We assume that auto-complete suggestions are most useful if they are not
much longer than the current input string $t$.
Therefore, the QCM only searches bins containing literals of length
$|t|$ to $|t| +\gamma$,
which reduces the cost of the sequential search. In this paper, we
use $\gamma = 10$.
When the search in residual bins completes, the
shortest result literals are returned 
as part of the
$k$ auto-complete suggestions. 

 
To ensure interactive response time,
we parallelize the QCM's sequential search in the residual bins,
utilizing $P$
parallel processes (threads).
Typically, $P$ would equal the number of available
cores on the \system server.
Each process searches one or more bins, and the QCM assigns work to processes in a way that balances load, with each process scanning an equal number of
literals.
Algorithm~\ref{algo:workload} shows the details of task assignment.

\subsection{Query Suggestion Module}
\label{subsec:qsm}

\begin{algorithm}[t]
\SetKwData{Left}{left}\SetKwData{This}{this}\SetKwData{Up}{up}
\SetKwFunction{Union}{Union}\SetKwFunction{FindCompress}{FindCompress}
\SetKwInOut{Input}{input}\SetKwInOut{Output}{output}

\Input{Query $q$, Predicate Set $PR$, Literal Bins to Search ${bins}'$, Number of Processes $P$}
\Output{Alternative Queries ${Q}'$}
\BlankLine

\For{each triple $tr$ in $q$}{
\For{each non-variable element $e$ in $tr$}{
\eIf{$e$ is a predicate}{
Lexica for term $S = Lemon.getLexica(e)$\;
\For{Each element $s$ in $S$}{
Predicate alternatives $pa.add$(FindPredicateAlternatives($s$, $PR$, $P$))\;
}

}
{
Literal alternatives $la.add$(FindLiteralAlternatives($e$, ${bins}'$, $P$))\;
}
}
}
SortBySimilarityScore($pa$)\;
SortBySimilarityScore($la$)\;
\For{For each alternative $a$ in $pa$}{
Construct a new query ${q}'$\;
Alternative queries for predicates $PQ.add({q}')$\;
}
\For{For each alternative $a$ in $la$}{
Construct a new query $q'$\;
Alternative queries for literals $LQ.add({q}')$\;
}
${Q}'.add($TopQueriesWithAnswer($PQ, k/2$)$)$\;
${Q}'.add($TopQueriesWithAnswer($LQ, k/2$)$)$\;
return ${Q}'$\;
\caption{Suggesting Alternative Query Terms}
\label{algo:alternatives}
\end{algorithm}

\begin{figure*}[t]
\centering
\includegraphics[height=6.5cm,width=2\columnwidth]{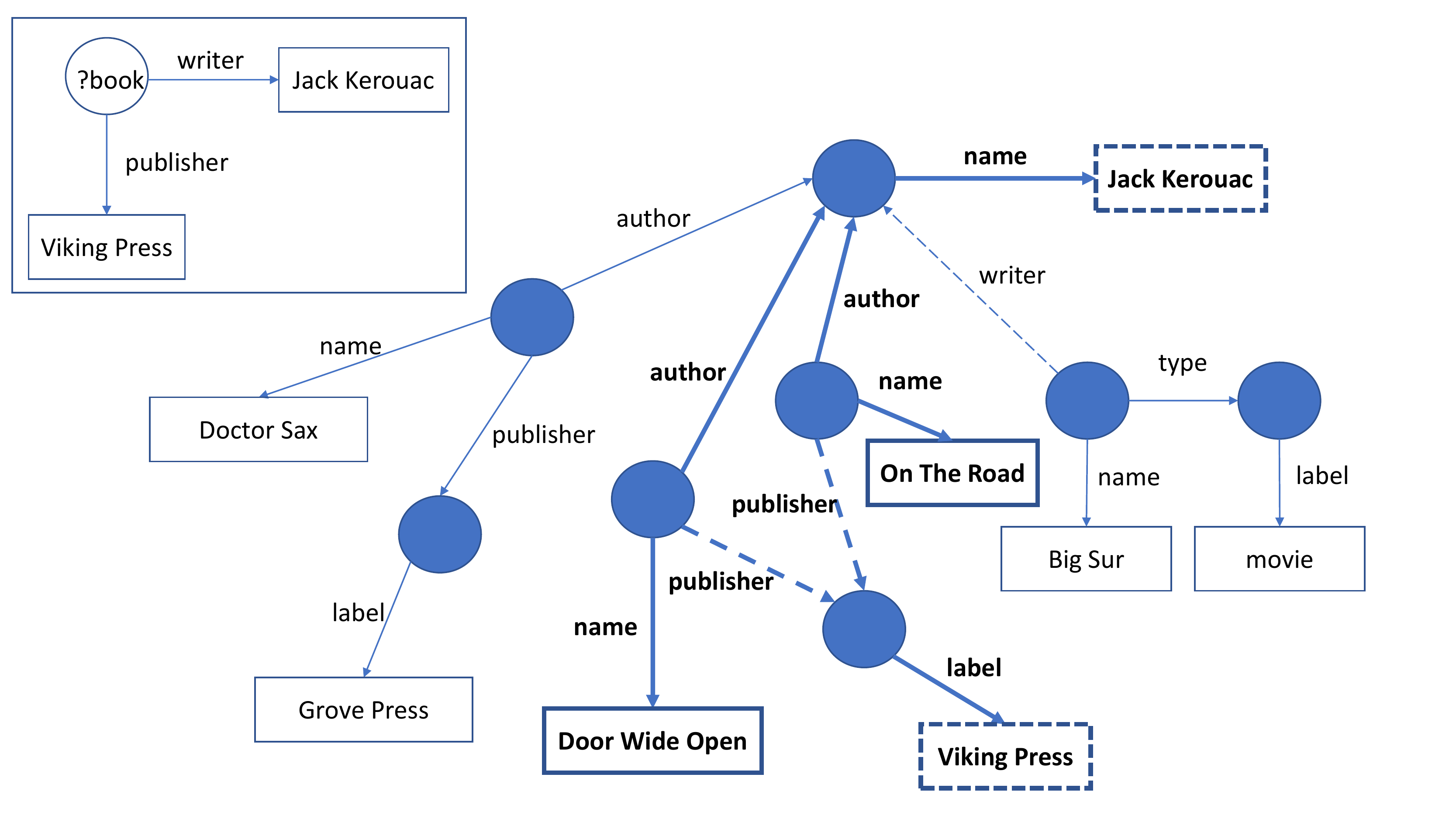}
\caption{Example query and the subgraph from the dataset that can be used to answer this query.}
\label{fig:relaxExample}
\end{figure*}

\begin{figure*}[t]
\centering
\includegraphics[height=6.5cm,width=2\columnwidth]{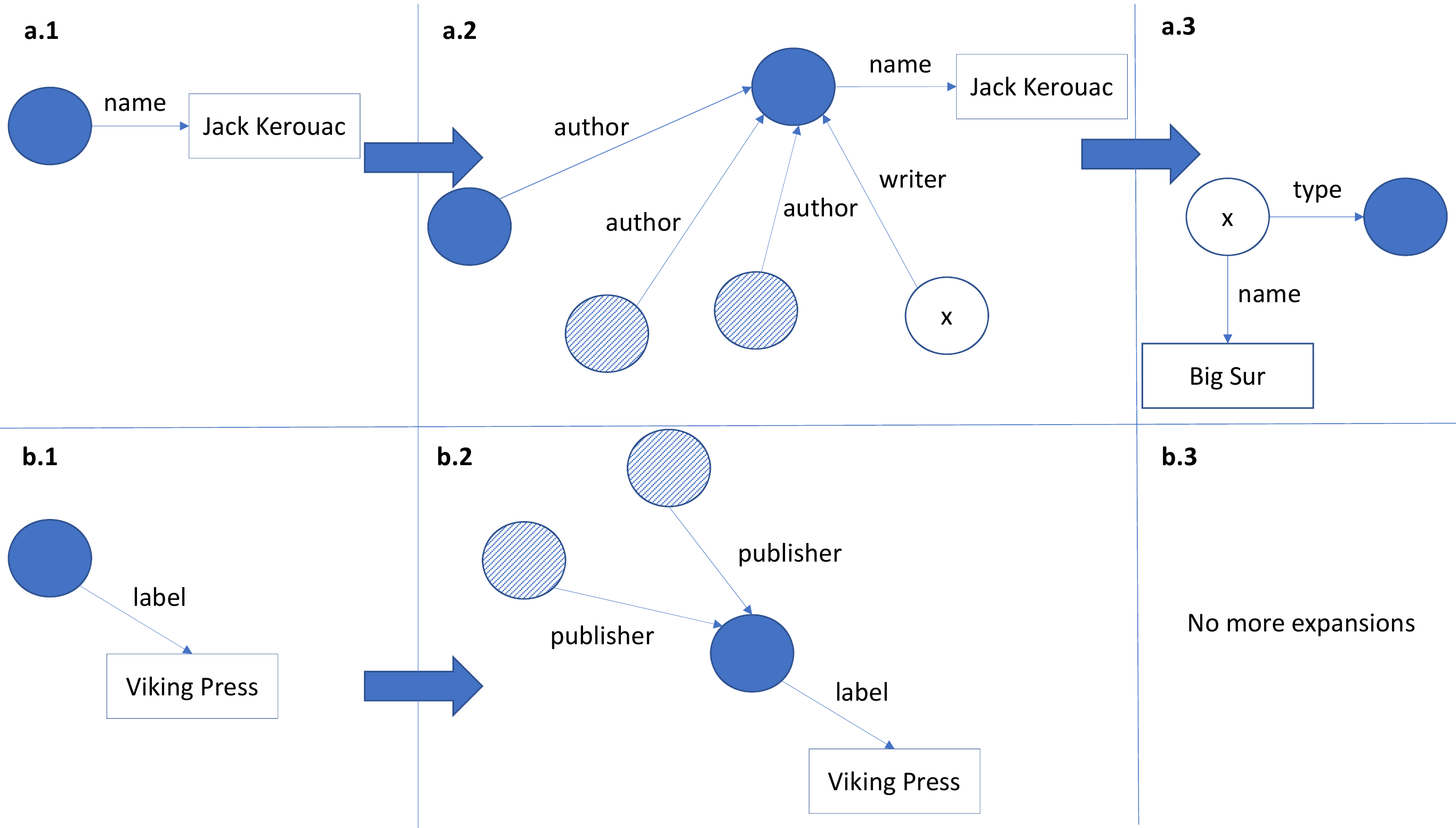}
\caption{The expansion steps in the relaxation process.}
\label{fig:relaxSteps}
\end{figure*}

The QSM suggests alternative queries that are \emph{semantically close} to
the query issued by the user.
The suggestions of the QSM are particularly important if the query issued
by the user returns no answers,
but they can be useful even if the query returns answers.
Defining semantic closeness is an interesting question.
In \system, the QSM suggests changes to the query in two directions:
(1)~Suggesting alternatives to the terms (predicates and literals) used in
the query, and
(2)~Relaxing the structure of the query.

\subsubsection{Alternative Query Terms}
\label{subsubsec:alternative}

Algorithm~\ref{algo:alternatives} shows how the QSM finds alternatives for
predicates and 
literals in the user  query.
The basic idea is to find predicates and literals in the data set that are
similar to the ones in the query or to their lexica. The lexica provides knowledge about how properties, classes and individuals are verbalized in natural language. For example, ``wife'' or ``husband'' can be verbalized by using ``spouse'' instead. 
The QSM examines the predicates and literals used in the triple patterns of the query one
at a time.
 For each predicate $p$, the QSM first finds the lexica for the
 predicate (line 4). We use the DBpedia Lemon Lexicon~\cite{Cimiano2014,Unger2013} to provide such lexica for the terms typed in by the user.
The QSM then finds alternative predicates in the dataset whose similarity
score with the
 original predicate $p$ or its lexica exceeds a similarity
 threshold $\theta$.
In \system, we use Jaro-Winkler (JW) similarity~\cite{Cohen2003} to
 calculate the similarity between strings.
JW similarity is based on the minimum number of single-character
transpositions required to change one string into the other,
while giving a more favorable score to strings that match from the beginning. This similarity measure outperforms other similarity measures in our context. In this paper, we use the $\theta = 0.7$.
For each literal $l$, the QSM considers the bins containing literals of
length in the 
range $[|l| - \alpha, |l| + \beta]$ (termed ${bins}'$ in
Algorithm~\ref{algo:alternatives}). A search operation over these bins is
conducted, similar to the search over bins in the QCM.
The difference is that the search to find alternative literals is based on
the JW similarity. 
All literals that have a similarity score $\geq \theta$ are considered to
be matches. 
We use 
the values
$\alpha = 2$ and $\beta = 3$.
The lists of alternative predicates and literals are sorted based on the
JW similarity score.
Similar to the QCM, the QSM can parallelize finding alternative terms among
$P$ processes. 

The alternative terms are sorted based on their similarity scores, and a new SPARQL query
is
constructed for each of the alternative predicates and literals found by
the QSM.
\system uses the federated query processor to execute the alternative queries 
and suggests the top queries that return answers.

\subsubsection{Relaxing Query Structure}


%
%
%
%
%

\begin{algorithm}[t]
\SetKwData{Left}{left}\SetKwData{This}{this}\SetKwData{Up}{up}
\SetKwFunction{Union}{Union}\SetKwFunction{FindCompress}{FindCompress}
\SetKwInOut{Input}{input}\SetKwInOut{Output}{output}

\Input{Query $q$}
\Output{Matching Graphs $G_{suggested}$}
\BlankLine
Literals in query $L = q$.extractLiterals()\;
\For{Each literal $l$ in $L$}{
  Seed group $seeds(l) = l \cup $ Top $k-1$ literals from $la(l)$\;
}
Start with empty graph $g$\;
\While{$g$ does not span terminals from all seed groups}{
Scan vertices using Dijkstra's bi-directional shortest path algorithm\;
Select a terminal $x$ not in $g$ that is closest to a vertex in $g$ (initially any literal from the query)\;
Add to $g$ the shortest path that connects $x$ with $g$\;
}
// There can be several $g$ subgraphs spanning \\// terminals if multiple paths with the same weight\\// cost exist\\
\For{Each $g$ found while connecting seeds}{
Construct subgraph $g'$ induced by $g$ in $G$\;
Construct minimum spanning tree(s) of $g'$\;
\While{There exist non-terminals of degree 1 from spanning tree(s)}{
remove non-terminals of degree 1 from this spanning tree\;
}
Add minimum spanning tree(s) to $G_{suggested}$\;
}
Return $G_{suggested}$\;
\caption{Relaxing Query Structure}
\label{algo:relaxation}
\end{algorithm}



%


If the structure of the graph pattern specified by the user in the query is different from the structure of the
queried dataset, the user will not find the desired answer,
even if the predicates and literals in the query match
the desired answer in the dataset.
Therefore, the QSM suggests changes to relax the structure of the query
(i.e., make it less constrained) based on
the structure of the dataset.

Figure~\ref{fig:relaxExample} shows a motivating example.
The query in this example is syntactically correct (top left box), and it
aims to find books by ``Jack Kerouac'' that were published by ``Viking
Press''. 
The figure shows part of the graph of the queried dataset.
The predicates and literals of the query can be found in the dataset, and
the matches are shown in the figure as dotted lines and rectangles.
The figure also shows two answers that satisfy the query requirements, and the path that connects them in bold
(``Door Wide Open'' and ``On the Road'').
These answers will not be found by the query as posed by the user since the
query structure does not match the structure of the data (the dotted
matches are not connected).
Relaxing the query structure can solve this problem by bringing
the structure of the query closer to the structure of the dataset.


In \system, we assume that it is easier for the user to identify correct
literals than to identify correct query structure.
Thus, we define the goal of query relaxation to be connecting
literals in the query
(or similar literals found by the JW similarity search)
through valid paths
in the graph of the dataset. Ideally, the paths should be short and the
algorithm should prefer paths that include the predicates
entered by the user as part of the query.
We observe that connecting the literals in the query can be formulated as a
\emph{Steiner tree} problem~\cite{Karp1972}, and that favoring paths that
include certain predicates can be achieved by modifying the weights on the edges of the graph.

The Steiner tree problem is defined as
follows. In any undirected graph $G = (V, E)$, where $V$ is the set of
vertices and $E$ is the set of edges, and each edge $e_{ij}$
connecting vertices $(i, j)$ has a weight $w_{ij}$, the Steiner
tree problem is finding a minimum weight tree that spans a subset of
terminal vertices (literals in our case) $T \subset V$.
If $T = V$, the problem is reduced to a minimum spanning tree problem. If
$|T| = 2$, the problem is reduced to a shortest path problem.
However, when
$2 < |T| < |V|$, finding a minimum weight tree is NP-Hard.

We associate a weight with each edge in the graph of the dataset.
These weights can be inferred by the algorithm and do not need to be
materialized.
For an edge representing a predicate that matches one of the predicates
in the query, or one of the predicates identified
by the process in Section~\ref{subsubsec:alternative}
as an alternative query term for a predicate in the query,
this weight is $w_{q}$.
For any other edge, the weight is  $w_{default} > w_{q}$.
Since the Steiner tree algorithm aims to find the tree with the minimum
overall weight,
assigning weights in this manner favors matching the predicates in the query
(or alternatives to these predicates) over simply finding a tree with a small 
number of edges.

Since finding the Steiner tree is an NP-hard problem, we need an
efficient approximate algorithm. 
Moreover, traditional Steiner tree algorithms, whether exact or
approximate,
require fast access to any vertex or edge in the
graph,
whereas in our case 
the graph exists on remote endpoints and can be accessed only through
SPARQL queries.
Our algorithm must 
minimize the number of such queries.
We describe next, (1) the literals to be connected via the Steiner tree
algorithm, and (2) the algorithm that we use to connect these literals.  


In the previous section, we described how we generate alternative
query terms for the 
literals in the query (line 9 in Algorithm~\ref{algo:alternatives}).
Each literal in the query and the alternative terms generated for it
form a \textit{group}, and we refer to the vertices representing
these
 literals in the RDF graph as the
\emph{seeds}
 of the
 QSM
for exploring the graph.
For example, ``Viking Press'', ``The Viking Press'', and ``The Viking'' are
all seeds
 in the same group.
The goal of our algorithm is to create a Steiner tree that connects
one literal from each group. It is not useful to connect multiple literals
from the same group since these literals are alternatives to each other and
not meant to be used together in the same query.

To connect the literals efficiently, our algorithm expands the graph
starting from the seeds until the groups are all connected, and it attempts
to minimize the number of vertices visited in this expansion.
We use a known Steiner tree approximation algorithm and we adapt it for our
use case~\cite{Hwang1992}. The details are presented in Algorithm~\ref{algo:relaxation}, and consist of the following two steps:

\textbf{1.~Connecting seeds:} 
The goal of this step is to find a tree, not necessarily minimal, that
connects all groups.
Initially, each seed is a \emph{candidate subgraph} of the RDF graph,
and the candidate
subgraphs are expanded using the \textit{bi-directional Dijkstra shortest path
algorithm}~\cite{Goldberg2005}.
In this algorithm,
seeds from different groups take turns in expansion rather than choosing a single source seed from which to start the expansion.
In practice, this approach visits (expands) fewer vertices than the regular Dijkstra shortest path algorithm,
which means fewer SPARQL queries.
The expansion continues until paths are found that connect seeds from all
groups.



In the expansion, each vertex $v$ in a candidate subgraph is expanded into a subgraph $subG$
defined as follows: 
(1)~$subG = \{(?s, ?p, ?o) | ?o = v\}$ if $v$ is a literal,
and (2)~$subG = \{(?s, ?p, ?o)| ?s = v \vee ?o = v\}$ if $v$ is a URI.
That is, if the vertex is a literal (initially, all vertices are literals), the subgraph is expanded by finding all triples that have this literal as an object since literals can
only be objects. Each of these triples introduces a new edge (the predicate) and vertex (the
subject) to the candidate subgraph. If a vertex is a URI, the subgraph is expanded by finding all triples that have this vertex as a subject or an object. As in the case of literal vertices, each of these triples introduces a new vertex to the candidate subgraph (the subject of the triple if the expanded vertex is the object, and the object if the expanded vertex is the subject). The edge connecting the new vertex to the expanded vertex is the predicate of the triple. These expansion steps are expressed as SPARQL queries executed on the endpoint of the dataset.

The algorithm expands candidate subgraphs according to the
bi-directional Dijkstra algorithm 
until it finds a shortest path that
connects two seeds from different groups. This path becomes the graph $g$
that will be used to find the
tree connecting all the groups.
The expansion of other candidate subgraphs continues according to the
bi-directional Dijkstra algorithm,
and whenever the expansion of a candidate subgraph results in connecting
to $g$ a seed from a group that is not yet part of $g$, the path that
connects this seed to $g$  is added to $g$.
The expansion stops when there is a set of connected seeds, one from each
group. 
Recall that we assign lower weights to the edges in the data
matching predicates
in the query or similar predicates. This  
guides the bi-directional Dijkstra algorithm towards
expanding paths that match query predicates first,
and consequently reduces the number of queries required
to find a tree that matches the query predicates.

We provide the expansion algorithm with a budget for the number of queries
that can be used. In order to remain within the budget, the expansion of
sibling vertices that are chosen for expansion does not start if
the number of siblings is larger than the remaining query budget. This
restriction discourages the expansion of vertices with a high degree
branching factor with the hope that this candidate subgraph's seed can be
reached by another seed from a different group.
We use a budget of 100 SPARQL queries for graph expansion, and we found
that this gives us good response time for query suggestion.
While expanding the candidate subgraphs, the results of the expansion are
memoized so that if a vertex is encountered more than once during
expansion, the results will be obtained from the memoized data structure
without issuing a new SPARQL query.

Figure~\ref{fig:relaxSteps} shows how the vertices in the example 
are expanded starting from
the seeds in the query. Common vertices between candidate subgraphs are
lightly shaded. All the edges have a cost of $w_{default}$ except for
``writer'' and ``publisher'', which have a cost of $w_{q}$. Therefore, ``writer'' is chosen to
be expanded in step a.3. However, this vertex will not be further expanded
because the expansion did not result in any common vertices with the
subgraph of the other literal in the query. Therefore, it is not possible that further expansion
will help finding a shorter path than the one already found. 

\textbf{2.~Constructing the minimum tree:} After the expansion step, we
construct a graph $G$ consisting of the union of all expansions.
For each $g$ found during expansion, we construct a subgraph $g'$ which is
the graph induced by $g$ in $G$. That is, $g'$ is a graph whose vertices
are the same as $g$ and whose edges are the edges in $G$ such that both
ends of the edge are vertices in $g$. 
Next, a minimum spanning tree is constructed for subgraph $g'$. 
Multiple minimum spanning trees may exist and be generated in this step.
Finally, all non-terminal vertices that have a degree of 1 are repeatedly deleted from
the minimum spanning tree(s) since they cannot be part of the Steiner tree. There could be
several Steiner trees with the same total edge weight.
Each tree is an alternative query
suggested to the user.
The approximation ratio of this algorithm is known to be $2 -2/s$~\cite{Hwang1992}, where $s$ is the number of seeds in the query.

\textbf{Performance:}
Unlike the QCM, which should have sub-second latency to provide
suggestions while the user types, the QSM can have a latency of a few
seconds. That is, after the user submits a query, she will see alternative,
complete, and syntactically correct suggested SPARQL queries after waiting a
few seconds. 
In querying LOD data using SPARQL, a query will likely have a small number
of literals (in our user study, the maximum number of literals in a query
was 3).
Our algorithm is fast enough for such problem sizes to guarantee a QSM response
time of less than
10 seconds on average.

\section{Evaluation}
\label{sec:evaluation}

\begin{sloppypar}
We evaluate \system along the following dimensions: 1.~A user study in
which participants answer questions using a natural language QA system and
\system (Section~\ref{subsec:user}).
2.~A quantitative comparison with recent natural language,
approximate query, and query-by-example systems
(Section~\ref{subsec:comparison}).
3.~Analyzing the response time of the QCM and QSM modules
(Section~\ref{subsec:performance}).
\end{sloppypar}


\system is implemented in Java. It runs as a web application over a web server.
The user interacts with \system through a web browser
as described in~\cite{El-Roby2016}.
We use a publicly available implementation of the suffix tree construction algorithm~\cite{Ukkonen1995},
FedX~\cite{Schwarte2011} as the federated query processor,
and the lemon lexicon for DBpedia\footnote{\url{http://github.com/ag-sc/lemon.dbpedia}}, which can be also used for other data sets.
We use DBpedia in all our experiments and we interact with it via its
SPARQL endpoint\footnote{\url{http://dbpedia.org/sparql}}.
DBpedia is a good evaluation dataset because it is
large and it is the central and most connected multi-domain dataset in the LOD cloud\footnote{\url{http://lod-cloud.net}}.
We run our experiments on a machine with
an 8-core Intel i7 CPU at 2.6 GHz and 8GB of memory.
The memory usage of \system to query DBpedia never exceeds 4GB.



\subsection{User Study} 
\label{subsec:user}

\subsubsection{User Study Setup}

The most important question related to \system is whether it actually helps users find answers in RDF datasets.
To answer this question, we conducted a user study in which users are presented with a set of questions they need to answer 
using both \system and QAKiS~\cite{Cabrio2012}, a natural language question answering system that performs well compared to the other natural language systems (see Section~\ref{subsec:comparison}).

The questions in our study are a subset of the query set from the
Schema-agnostic Queries Semantic Web Challenge~\cite{SAQ}. These queries
are questions over DBpedia derived from the Question Answering over Linked Data (QALD) competition\footnote{\url{http://qald.sebastianwalter.org}}.
We started with 35 questions and divided them into three difficulty categories (easy, medium, and difficult). Each of the authors of this paper independently labeled each question as easy, medium, or difficult.
Out of the 35 questions, the authors all agreed on the difficulty level of  27 questions,
and we used these questions in the  user 
study. These queries are available in Appendix~\ref{app:querySet}.

\begin{sloppypar}
We recruited 16 participants who have a computer science background but are not familiar with RDF or SPARQL. 
Each participant was given 10 questions (4 easy, 3 medium, and 3 difficult).
The questions were randomly assigned to participants per category.
We asked  the participants to find answers to all the questions using both \system and 
QAKiS. Since they are fundamentally different in the way they are used, using one system to find an answer should have minimal effect on how the other system is used.
However, we alternated the system the user used first for every question.
For example, if the participant answers one question using \system first
then QAKiS, the next question is answered using QAKiS first then \system.
One question from the easy category was used in a tutorial prior to the study to demonstrate the two systems to the users (the same question for all participants).
During the study, the first question a participant tried (from the easy category) was used as a warm-up question to familiarize the user with the two systems. 
The data we collected for this first question is dropped from the results.
We used screen recording to capture the sessions of all participants.
\end{sloppypar}

\subsubsection{Quantitative Results}

\begin{figure}[t]
\centering
\includegraphics[height=3.5cm,width=0.9\columnwidth]{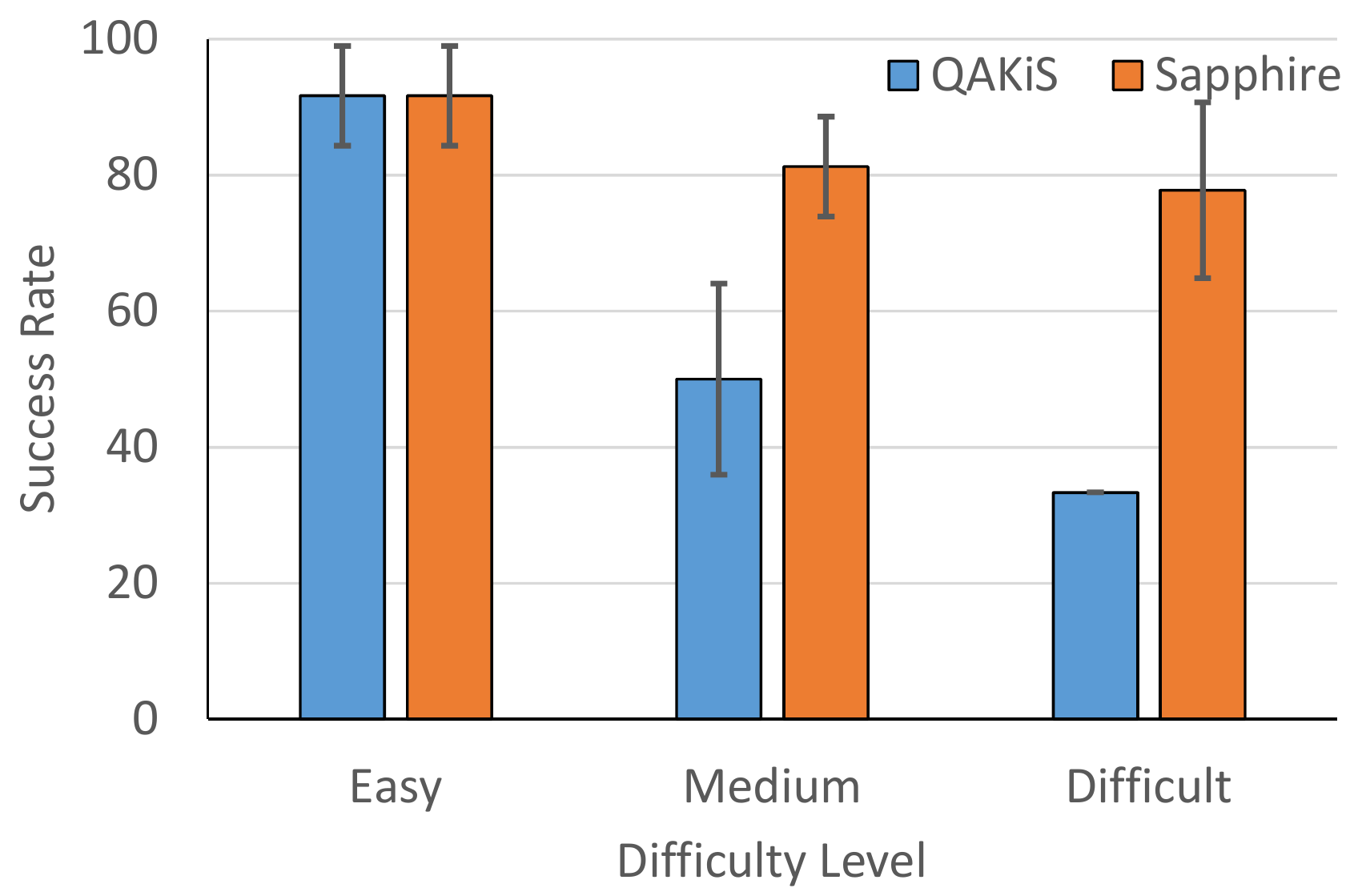}
\caption{Success rate of answering questions.}
\label{fig:findingAnswers}
\end{figure}

\begin{figure}[t]
\centering
\includegraphics[height=3.5cm,width=0.9\columnwidth]{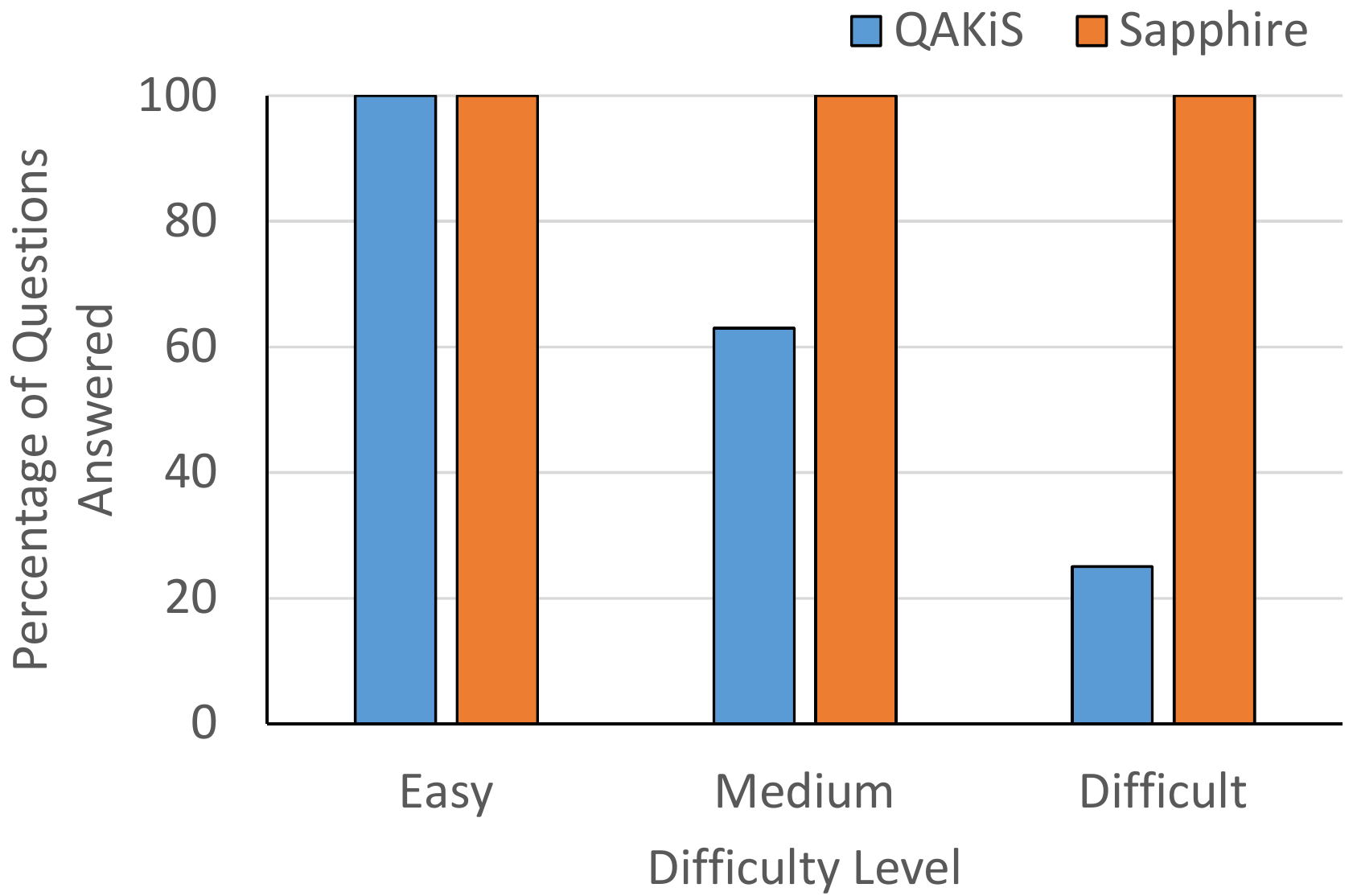}
\caption{Percentage of questions answered by at least one participant.}
\label{fig:findingAnswersPercentage}
\end{figure}

\begin{figure}[t]
\centering
\includegraphics[height=3.5cm,width=0.9\columnwidth]{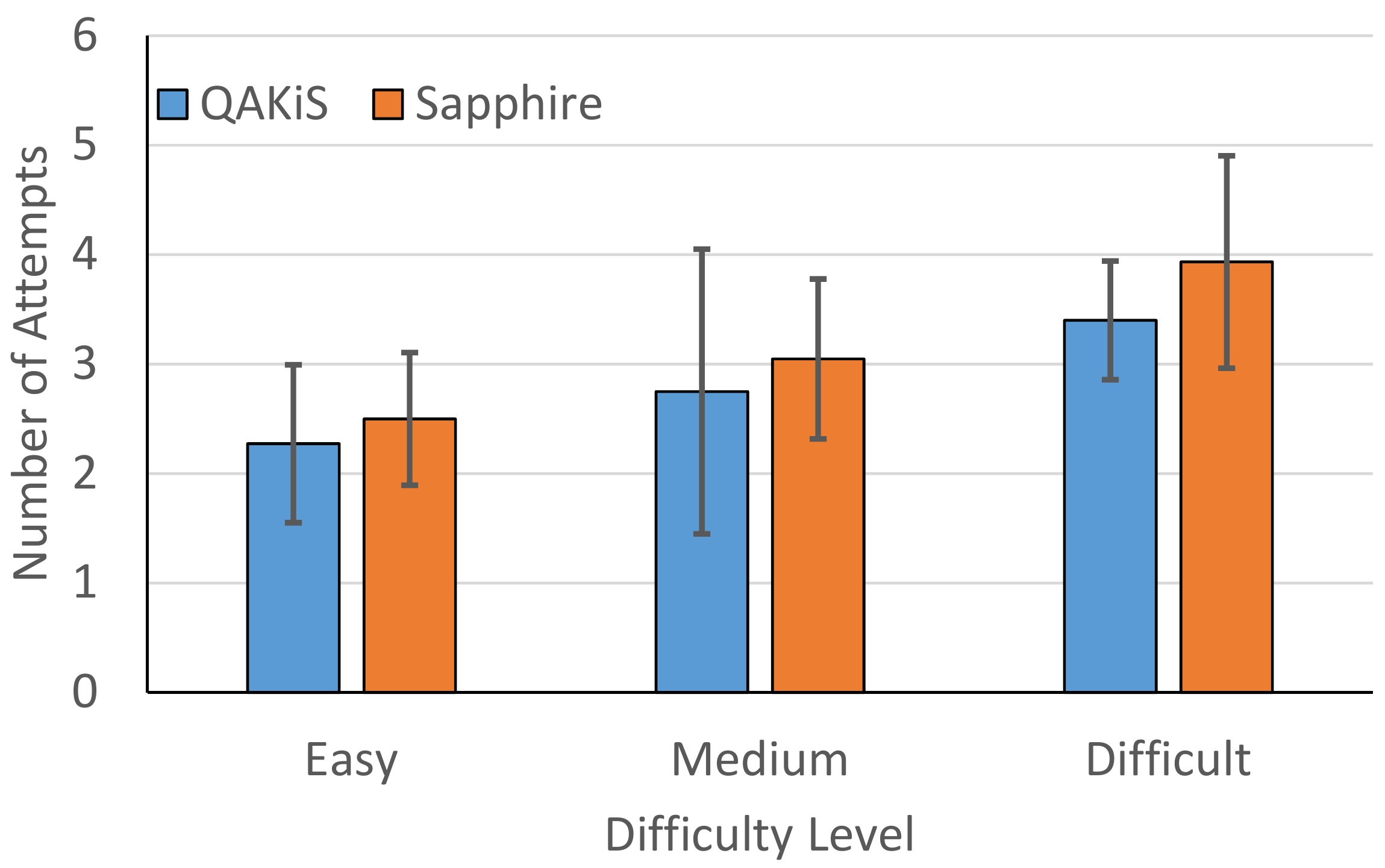}
\caption{Average number of attempts before finding an answer.}
\label{fig:attempts}
\end{figure}

\begin{figure}[t]
\centering
\includegraphics[height=3.5cm,width=0.9\columnwidth]{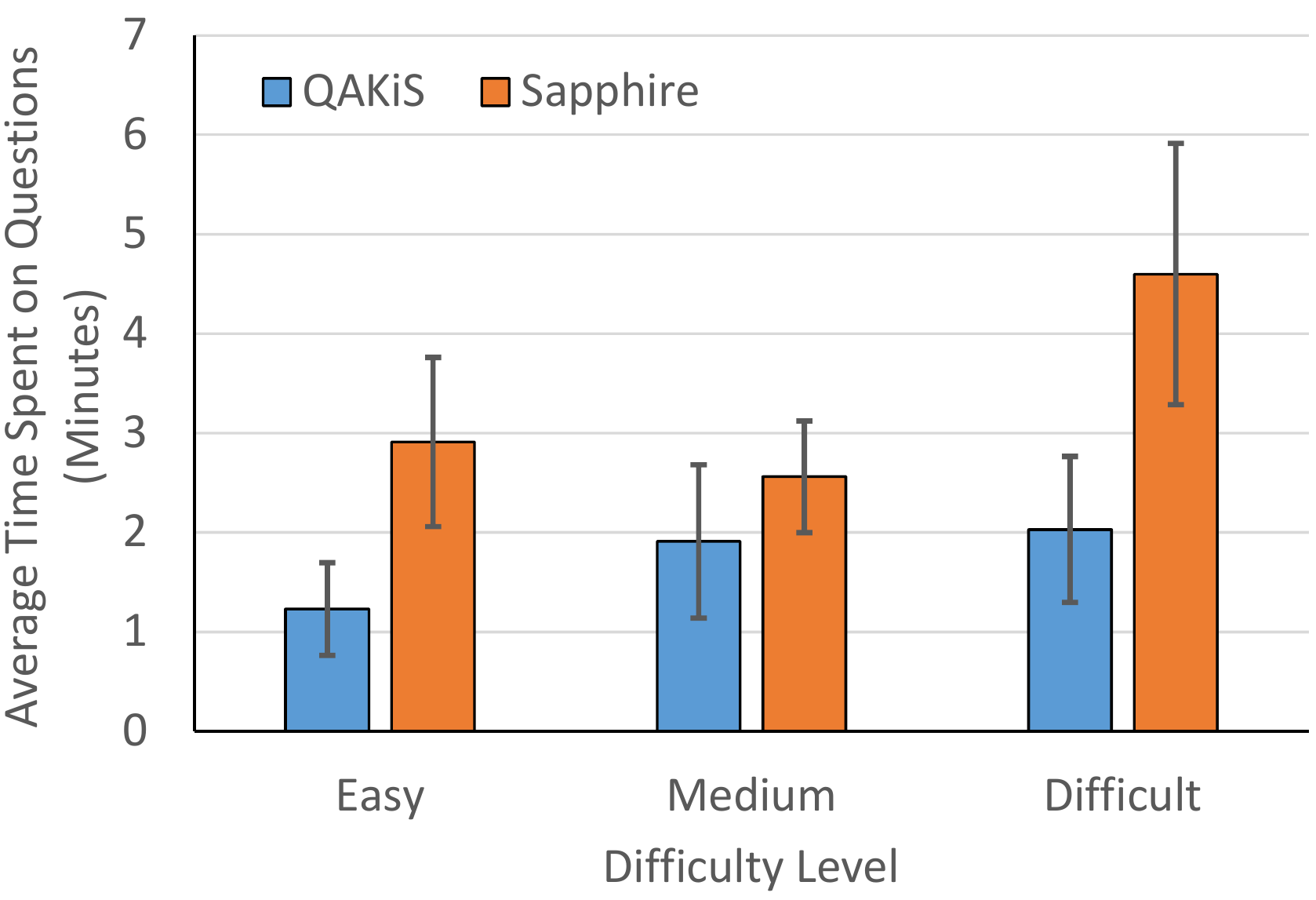}
\caption{Average time spent on answered queries.}
\label{fig:queryTime}
\end{figure}


We first investigate whether \system helped users find answers to their assigned questions, and how it compares to QAKiS.
A total of 48 questions in each category was given to the participants in this study 
(16 participants $\times$ 3 questions per category, excluding the first warm-up question).
We first study the success rate in answering these questions. That is, of the questions given to a user, what fraction was answered correctly?
Figure~\ref{fig:findingAnswers} shows the success rate of finding answers for the 48 questions in each category using \system and QAKiS. 
The bars in the figure show the average success rate, averaged over the 16 participants.
 We also report the 95\% confidence interval to demonstrate that the
 findings are consistent among all participants. Whenever we observed a
 noticeable difference between \system and QAKiS (in all experiments), we
 calculated the p-value and found it to be less than the significance level (0.05), which indicates that these differences are statistically significant. The figure shows that
 \system is superior to QAKiS in the medium and difficult categories,
 while both systems perform the same for the easy category. Participants
 found answers for over 80\% of the medium difficulty questions using
 \system, compared to around 50\% using QAKiS. The gap widens for the
 difficult category, where participants
 answered almost 80\% of the questions using \system
and only 35\% using QAKiS. 

The success rate does not tell the full story since some questions are easier than others and some users are better at answering questions than others, regardless of the difficulty category or the system used.
Another way to compare the two systems is to see, for every question, whether that question was answered by any participant.
Figure~\ref{fig:findingAnswersPercentage} shows the percentage of questions answered
by at least one participant using both systems.
The figure shows that every question was answered by at least one participant using \system, while QAKiS could find answers for only 63\% of the questions in the medium category and 30\% in the difficult category.

Figure~\ref{fig:attempts} shows the average number of attempts the participants went through
before finding an answer in each category.
An attempt is counted when a participant clicks ``Run'' to issue a query.
\system requires slightly more attempts than QAKiS, but the numbers are comparable and not statistically significant (p-value $>$ 0.05).
This demonstrates that \system is not overly  difficult to use despite the need to describe a query in a structured format. 
Note that attempts are counted only for the questions that were answered correctly.
Participants gave up on finding an answer for a question after 3 to 4 attempts when using QAKiS, and after 3 to 5 attempts when using \system.

\system does require more time to use than QAKiS, as demonstrated in Figure~\ref{fig:queryTime}, 
which shows the time the participants spent on questions from each category. The figure only shows the time spent on questions that were answered successfully.
The figure shows that participants spent more time using \system than QAKiS for all
difficulty categories. 
This is expected due to the fundamentally different approach of describing
a question in \system. A participant spends more time to describe the
question as a set of triple patterns, and to examine \system's suggestions
and choose from them. This additional effort is justified by \system's
ability to find answers to more questions.

In summary, the user study shows that \system is more effective than QAKiS at answering medium and difficult questions. The cost of this effectiveness is more time spent in answering the questions.
  

\begin{table*}
\small
 \centering
 \begin{tabular}{| c | c | c | c | c | c | c | c | c | c | c |}
      \hline\hline
 & $\#pro$ & $\%$ & $\#ri$ & $\#par$ & $R$ & $R^{*}$ & $P$ & $P^{*}$ & $F_{1}$ & ${F_{1}}^{*}$\\ [0.5ex]
 \hline
 Xser~\cite{Xu2014} & 42 & 84\% & 26 & 7 & 0.52 & 0.66 & 0.62 & 0.79 & 0.57 & 0.72 \\\hline
 APEQ~\cite{Unger2015} & 26 & 52\% & 8 & 5 & 0.16 & 0.26 & 0.31 & 0.50 & 0.21 & 0.34 \\\hline
 QAnswer~\cite{Ruseti2015} & 37 & 74\% & 9 & 4 & 0.18 & 0.26 & 0.24 & 0.35 & 0.21 & 0.30 \\\hline
 SemGraphQA~\cite{Beaumont2015} & 31 & 62\% & 7 & 3 & 0.14 & 0.20 & 0.23 & 0.32 & 0.17 & 0.25 \\\hline
 YodaQA~\cite{Unger2015} & 33 & 40\% & 8 & 2 & 0.16 & 0.20 & 0.24 & 0.30 & 0.19 & 0.24 \\\hline
 \hline
 QAKiS~\cite{Cabrio2012} & 40 & 80\% & 14 & 9 & 0.28 & 0.46 & 0.35 & 0.58 & 0.31 & 0.51 \\\hline
 KBQA~\cite{Cui2017} & 8 & 16\% & 8 & 0 & 0.16 & 0.16 & \textbf{1.0} & \textbf{1.0} & 0.28 & 0.28 \\\hline
 $S^{4}$~\cite{Zheng2016} & 26 & 52\% & 16 & 5 & 0.32 & 0.42 & 0.62 & 0.81 & 0.42 & 0.55 \\\hline
 SPARQLByE~\cite{Diaz2016} & 7 & 14\% & 4 & 0 & 0.08 & 0.08 & 0.57 & 0.57 & 0.14 & 0.14 \\\hline
 \system & \textbf{43} & \textbf{86\%} & 43 & 0 & \textbf{0.86} & \textbf{0.86} & \textbf{1.0} & \textbf{1.0} & \textbf{0.92} & \textbf{0.92} \\\hline
    \end{tabular}   
     \caption{Comparing systems using questions from QALD-5.}
     \label{tab:qald5}
\end{table*}

\subsubsection{Qualitative Results}

After each session, we surveyed the participants about their experience using \system and how it compares to QAKiS. The comments we received are consistent across participants: At first, they find it difficult to express the question using triple patterns (due to the lack of experience in RDF) but are still able to answer the questions.
However, when they get used to this style of querying, \system becomes much easier to use.
They also agree that \system is much more helpful than QAKiS in answering more difficult questions.

\begin{sloppypar}
Another observation we make from viewing the recorded sessions is that different participants answering the same question sometimes take different approaches and use different terms, but end up with the same SPARQL query.
In other cases, different participants end up with different queries to find the same answer.
For example, some participants rank results by a condition and select the correct answers while others include the condition in the triple patterns of the query.
This
demonstrates the flexibility and effectiveness of \system. 
\end{sloppypar}

For another qualitative perspective on \system, we recruited two SPARQL experts,
one with no experience in querying DBpedia and the other with  three years of experience.
The two participants were asked to write SPARQL queries to find answers to the 48
questions used in the user study, with and without \system.
Without \system, i.e., interacting directly with the SPARQL endpoint of
DBpedia, the first participant was unable to answer any of the questions because
he does not know how the DBpedia URIs are represented
and what kind of vocabulary is used in it.
When using \system, he was able to find answers to most questions.
The participant with three years experience in DBpedia answered most questions.
\system did help him answer the questions he failed to find answers for using DBpedia's SPARQL endpoint.
 Both experts agreed on \system's value in helping users to write SPARQL queries against data sources they are less familiar with and expressed interest in using \system for their future projects.

\subsection{Comparison to Other Systems}
\label{subsec:comparison}

In this section, we compare \system to other state of the art systems for
querying RDF data.
We compare to the systems participating in the 
QALD-5 competition~\cite{Unger2015}.
In addition, we also compare to
(a)~QAKiS, which is used in our user study,
(b)~the more recent natural language QA system KBQA~\cite{Cui2017},
(c)~the recent approximate query matching system $S^{4}$~\cite{Zheng2016},
and 
(d)~the recent query-by-example system SPARQLByE~\cite{Diaz2016}.
We use the questions from 
QALD-5 (50 questions) and the performance measures used in
this competition.
We copy the performance numbers of
the systems that participated in 
QALD-5 and of KBQA from~\cite{Cui2017}.
We obtain performance numbers ourselves for QAKiS and SPARQLByE,
both of which are publicly available.
We also obtained performance numbers ourselves for $S^{4}$,
which we implemented.

QAKiS is a natural language QA system, and we allow up to 3 attempts for
each question. In these attempts, we do not change the query terms using
our knowledge of the vocabulary.
For example, the question ``What is the revenue of IBM?'' can be
paraphrased in a different attempt as ``IBM's revenue'', but we would not
change it to ``IBM's income''. 

$S^{4}$ constructs a summary graph of the data in on offline step,
and accepts SPARQL queries that it rewrites to match the
structure of the data based on the summary graph.
$S^{4}$ expects the predicates and literals to be correct, so we 
use Sapphire to help us find predicates and literals that exist in DBpedia
when constructing the SPARQL query for $S^{4}$.
We compose the SPARQL query for $S^{4}$ based on the question in QALD-5,
restricting ourselves to the terms in the question.
$S^{4}$ rewrites the query and we execute the rewritten query using FedX.

SPARQLByE requires the user to  provide example answers. The system attempts to  learn the commonalities between these answers and capture them in a SPARQL query. The answers of this SPARQL query are presented to the user as additional candidate answers, and the user can mark them as correct or incorrect.
SPARQLByE requires at least two sample answers so we use it for questions that have three answers or more in their gold standard result.
We present two answers from the  gold standard result as inputs to SPARQLByE, and we 
provide feedback to the system 
until it finds the correct query or cannot learn any more (i.e., cannot modify the query).

When using \system, we only use terms from the question to enter the query,
as we did with other systems.
We then use \system's suggestions to complete and modify the query until an answer is found. 
We do not use our knowledge of the vocabulary to change the terms or query structure.

The systems are evaluated using the following performance measures~\cite{Unger2015,Cui2017}:
1.~The number of questions that are processed and for which answers are found ($\#pro$).
2.~The number of questions whose answers are correct ($\#ri$)
3.~The number of questions whose answers are partially correct ($\#par$).
In addition, the following recall and precision measures are computed,
where $\#total$ is the total number of questions in the question set:
 \textit{Recall}  defined as $R = \frac{\#ri}{\#total}$,
\textit{partial recall} defined as $R^{*} = \frac{\#ri + \#par}{\#total}$,
\textit{precision} defined as $P = \frac{\#ri}{\#pro}$,
\textit{partial precision} defined as $P^{*} = \frac{\#ri + \#par}{\#pro}$,
$F_{1}$  defined as  $2. \frac{P.R}{P + R}$, and
${F_{1}}^{*}$ defined as $2. \frac{P^{*}.R^{*}}{P^{*} + R^{*}}$.

Table~\ref{tab:qald5} shows the performance of the different systems.
The table shows that \system outperforms all other systems on all measures.
Natural language QA systems suffer from low precision
due to the challenge of inferring the structure and terms of a SPARQL query from
the
natural language formulation of the question.
This challenge is not faced by \system, which helps the user to directly construct SPARQL queries. Therefore,  \system has a precision of 1.0 for the questions it is able to answer.
Among the natural language systems, KBQA has precision of 1.0  
like \system, but it has much lower
recall. This is because KBQA focuses only on
factoid questions. If only the
factoid questions are considered, KBQA achieves a recall of 0.67,
still lower than \system.
$S^{4}$, while lower in performance than \system, performs better than other
systems.
SPARQLByE has much lower recall than other systems because it cannot answer most of the questions. 

The table justifies our choice of QAKiS as a representative QA system in our user study.
Other than Xser and $S^{4}$, QAKiS is the best performing system after 
\system in terms of recall and F-measure. 
Xser is not publicly available.
$S^{4}$ requires exact knowledge of the literals and URIs in
the queried dataset, which we deem too difficult for a user study.



\subsection{Sapphire Response Time}
\label{subsec:performance}

%

\subsubsection{QCM}
It is important for the QCM to provide auto-complete suggestions with very low response time in order to guarantee an interactive experience for users.
We measure the response time of the QCM in the user study.
Two components contribute to the response time of the QCM: the lookup in the suffix tree, and the sequential search in the bins of literals.
We have found that the total response time of these two components is, on average, 0.16 seconds when including 40K significant literals in the suffix tree and using 8 cores for the sequential search in the residual bins. This response time is low enough to provide a  good interactive experience.

We now study the two components of this response time in more detail. 
We have found that a lookup operation in the suffix tree takes approximately 0.25 milliseconds, regardless of the number of literals that are indexed.
This response time is certainly low enough for an interactive user experience.
Recall that matches in the suffix tree are returned immediately to the user before the search in the bins of literals begins.
Thus, having a hit (match) in the suffix tree greatly enhances the interactive experience, since the user sees auto-complete suggestions very quickly.
Even if these suggestions are not chosen by the user, they still give an impression of a responsive system.
Therefore, a higher hit ratio in the suffix tree is better for interactive response of the QCM.
The hit ratio (fraction of query terms for which a match is found in the suffix tree) depends on the number of literals included in the suffix tree. Our experiments show that even with only 40K literals in the suffix tree, we achieve a hit ratio of 50\%.


The second component of the QCM response time is the sequential search in the literal bins.
Recall that the bins to be searched are filtered based on the length of the term entered by the user.
We have found that, on average, this filtering eliminates 46\% of the literals to be searched.
The search in the residual bins takes 0.6 seconds when using 1 core, and 0.16 seconds when using 8 cores.
The takeaway of this experiment is that 
the QCM can provide interactive response time by utilizing more cores.

\subsubsection{QSM}



The logs of our user study indicate that participants used the suggestions
of the  QSM in over 90\% of the questions.
Users utilized
alternative predicates in 28\% of the questions,
alternative literals in 17\% of the questions,
and relaxed query structure in 67\% of the questions.
This demonstrates the crucial role the QSM plays in guiding the user towards correctly describing her questions. The QSM spends around 10 seconds on average before returning suggestions to the user. This is acceptable since the QSM does not interact with the user while she is typing. Instead, the user waits for suggestions from the QSM, and a 10 second wait is reasonable.

\section{Conclusion}
\label{sec:conclusion}
In this paper, we introduced \system, a tool that helps users construct SPARQL queries 
that find the answers they need in RDF datasets.
\system caches data from the datasets to be queried and uses this cached data to suggest
completions for SPARQL queries as the user is entering them,
and modifications to these queries after they are executed.
We have shown \system to be effective at helping users with no prior knowledge of the queried datasets answer complex questions that other systems fail to answer.
As such, \system is a valuable tool for 
querying the LOD cloud.

\balance
\bibliographystyle{abbrv}
\bibliography{uw-ethesis}

 \begin{appendix}
 
\section{Initialization Queries}
\label{app:bootstrappingQueries}

This appendix presents the SPARQL queries used in initializing \system.
SPARQL queries are typically provided with limited resources by the remote
endpoints.
A long-running query that is expected to consume a lot of resources may
 be rejected by the remote endpoint. If the query is accepted, it will
likely time out.
Therefore, the initialization queries of \system are broken down into
multiple 
queries that are less resource-intensive and therefore less likely to
time out. These queries are as follows.

1. Finding predicates sorted by their frequency (not a resource-intensive query):
\begin{verbatim}
Q1) SELECT DISTINCT ?p (COUNT(*) AS ?frequency)
WHERE {
?s ?p ?o 
}
GROUP BY ?p
ORDER BY DESC(?frequency)
\end{verbatim}

2. Finding literals and most significant literals:
The queries used to find literals need to be carefully structured to
minimize their execution time and the chances of timing out. 
The key to achieving this goal is
increasing the selectivity of the query.
We focus on two common characteristics of RDF data that are relevant to
\system:
1.~Entities are associated with RDF types or schema classes.
2.~Literals of interest in \system are associated with a limited set of predicates.

Some datasets are well-structured and have a hierarchy of RDF schema
classes, with each entity in the dataset belonging to a class.
This is the case for most of the datasets that we encountered on the LOD cloud. 
We can exploit this characteristic  by restricting the retrieval of literals to part of the
class hierarchy.
The following query finds all classes and their subclasses in a dataset:
\begin{verbatim}
Q2) PREFIX rdfs: <http://www.w3.org/2000/01/rdf-schema#>
PREFIX owl: <http://www.w3.org/2002/07/owl#>
SELECT DISTINCT ?class ?subclass
WHERE{
?class a owl:Class.
?class rdfs:subClassOf ?subclass
}
\end{verbatim}

For  datasets that do not have an RDF schema class hierarchy, we can
exploit the most used property in the LOD\footnote{\url{http://stats.lod2.eu/properties}} (RDF types) in the dataset.
The following query is used to find all types in the dataset sorted by their frequency:
\begin{verbatim}
Q3) SELECT DISTINCT ?o (COUNT(?s) AS ?frequency) 
WHERE{
?s a ?o.
}
GROUP BY ?o
ORDER BY DESC(?frequency)
\end{verbatim}

In both cases, the following query is used to find predicates sorted by the
number of associations to literals:
\begin{verbatim}
Q4) SELECT DISTINCT ?p (COUNT(?o) AS ?frequency) 
WHERE{
?s ?p ?o.
Filter (isliteral(?o))
}
GROUP BY ?p
ORDER BY DESC(?frequency)
\end{verbatim}

The top $k$ of these predicates are filtered based on whether they satisfy
the filtering conditions on the language of the literals they are
associated with and the length of these literals.
This filtering is done by issuing the following query multiple times, once
for each predicate. The placeholder \verb+$PREDICATE$+ is replaced with
the  current predicate being queried:
\begin{verbatim}
Q5) SELECT DISTINCT ?o
WHERE{
?s $PREDICATE$ ?o.
Filter (isliteral(?o) && lang(?o) = 'en' && 
strlen(str(?o)) < 80)
}
LIMIT 1
\end{verbatim}

After issuing these queries to retrieve and filter predicates, if the
dataset uses RDF schema classes, 
\system 
constructs the tree representing the class hierarchy.
Starting from the root of this tree, the following query is issued to find
if literals associated with entities of a certain class (type) \verb+$TYPE$+ with a
predicate \verb+$PREDICATE$+ can be found.
This query is issued iteratively, iterating over all classes and predicates:
\begin{verbatim}
Q6) SELECT DISTINCT ?o 
WHERE{
?s a $TYPE$.
?s $PREDICATE$ ?o.
Filter (isliteral(?o) && lang(?o) = 'en' && 
strlen(str(?o)) < 80).
}
\end{verbatim}

If a query on the class \verb+$TYPE$+ times out, queries over subclasses of
this class are issued. If the query succeeds and returns an answer,
then issuing the same queries over the subclasses is redundant.

In the case of datasets that do not use an RDF schema class hierarchy, we
need a different way to reduce query result size. For this, we use
\verb+LIMIT+
and \verb+OFFSET+. Specifically, we issue the following query multiple
times, iterating over \verb+$TYPE$+ and \verb+$PREDICATE$+, and using
\verb+LIMIT+ and \verb+OFFSET+ to paginate the answers so that the query
does not time out:

\begin{verbatim}
Q7) SELECT DISTINCT ?o 
WHERE{
?s a $TYPE$.
?s $PREDICATE$ ?o.
Filter (isliteral(?o) && lang(?o) = 'en' && 
strlen(str(?o)) < 80).
}
LIMIT $LIMIT$
OFFSET $OFFSET$
\end{verbatim}

Finally, we need to find the most significant literals. The following query
template is used for this, and it is issued iteratively similar to  Q7:
\begin{verbatim}
Q8) SELECT DISTINCT ?o (COUNT(?subject) AS ?frequency) 
WHERE{
?s a $TYPE$.
?subject ?p ?s.
?s $PREDICATE$ ?o.
FILTER(lang(?o) = 'en' && strlen(str(?o)) < 80)
}
GROUP BY ?o
ORDER BY DESC(?frequency)
LIMIT $LIMIT$
OFFSET $OFFSET$
\end{verbatim}

Recall that \verb+$PREDICATE$+ is associated with literals. Therefore, the literal filter is not added and only the filters on language and length are used. 

Much of the complexity of the above queries is to avoid timeouts at the
remote endpoints. This is important when using \system in a
\textit{federated} architecture. Recall that \system can also be used in a
\textit{warehousing} architecture, where all the datasets are stored
locally on the same server as \system.
In the warehousing architecture, no limitations are placed on querying the
dataset, e.g., no resource constraints and no timeouts.
This makes finding literals much simpler since we can issue long-running
SPARQL queries without worrying about timeouts.

Specifically, the following query can be used to find literals filtered by
length and language in the warehousing architecture (\verb+$LIMIT$+ and
\verb+$OFFSET$+ can still be used to restrict the number
of results returned, if needed):
\begin{verbatim}
Q9) SELECT DISTINCT ?o 
WHERE{
?s ?p ?o.
FILTER(isliteral(?o) && lang(?o) = 'en' &&
strlen(str(?o)) < 80)
}
GROUP BY ?o
LIMIT $LIMIT$
OFFSET $OFFSET$
\end{verbatim}

The following query finds the most significant literals in the warehousing
architecture if there are no timeout constraints (again, with \verb+$LIMIT$+ and
\verb+$OFFSET$+ if needed): 
\begin{verbatim}
Q10) SELECT DISTINCT ?o (COUNT(?s1) AS ?frequency) 
WHERE{
?s1 ?p ?s2.
?s2 ?p2 ?o.
FILTER(isliteral(?o) && lang(?o) = 'en' &&
strlen(str(?o)) < 80)
}
GROUP BY ?o
ORDER BY DESC(?frequency)
LIMIT $LIMIT$
OFFSET $OFFSET$
\end{verbatim}

\section{Queries Used for User Study}
\label{app:querySet}

\subsection{Easy Queries}
\begin{enumerate}
\item Country in which the Ganges starts
\item John F. Kennedy's vice president
\item Time zone of Salt Lake City
\item Tom Hanks's wife
\item Children of Margaret Thatcher
\item Currency of the Czech Republic
\item Designer of the Brooklyn Bridge
\item Wife of U.S. president Abraham Lincoln
\item Creator of Wikipedia
\item Depth of lake Placid
\end{enumerate}

\subsection{Medium Queries}
\begin{enumerate}
\item Instruments played by Cat Stevens
\item Parents of the wife of Juan Carlos I
\item U.S. state in which Fort Knox is located
\item Person who is called Frank The Tank
\item Birthdays of all actors of the television show Charmed
\item Country in which the Limerick Lake is located
\item Person to which Robert F. Kennedy's daughter is married
\item Number of people living in the capital of Australia
\end{enumerate}

\subsection{Difficult Queries}
\begin{enumerate}
\item Chess players who died in the same place they were born in
\item Books by William Goldman with more than 300 pages
\item Books by Jack Kerouac which were published by Viking Press
\item Films directed by Steven Spielberg with a budget of at least \$80 million
\item Most populous city in Australia
\item Films starring Clint Eastwood direct by himself
\item Presidents born in 1945
\item Find each company that works in both the aerospace and medicine industries
\item Number of inhabitants of the most populous city in Canada
\end{enumerate}

 \end{appendix}


\end{document}